\documentclass[aip,jcp,amsmath,amssymb,reprint]{revtex4-1}
\usepackage{graphicx}% Include figure files
\usepackage{dcolumn}% Align table columns on decimal point
\usepackage{bm}% bold math
\usepackage{mathrsfs}
\usepackage{amsmath, amssymb, amsthm, amsfonts}
\usepackage[dvipsnames]{xcolor}
\usepackage{ulem}
\usepackage{caption}
\usepackage[textfont=rm]{subcaption}
\begin{document}
\title{Identifying structural signature of dynamical heterogeneity via the local softness parameter}
\author{Mohit Sharma}
\address{\textit{Polymer Science and Engineering Division, CSIR-National Chemical Laboratory, Pune-411008, India}}

\author{Manoj Kumar Nandi}
\affiliation{\textit{Department of Engineering,
University of Campania ``Luigi Vanvitelli"
81031 Aversa (Caserta), Italy}}

\author{Sarika Maitra Bhattacharyya}
\email{mb.sarika@ncl.res.in}
\address{\textit{Polymer Science and Engineering Division, CSIR-National Chemical Laboratory, Pune-411008, India}}
\affiliation{\textit{Academy of Scientific and Innovative Research (AcSIR), Ghaziabad 201002, India}}
%\date{May 2021}

\begin{abstract}
 In this work, we study the relationship between the softness of a mean-field caging potential and dynamics at the local level. We first describe the local softness, which shows a distribution, thus identifying structural heterogeneity. We show that the lifetime of the softness parameter is connected to the lifetime of the well-known cage structure in supercooled liquids. Finally, our theory predicts that the local softness and the local dynamics is causal below the onset temperature where there is a decoupling between the short and long time dynamics, thus allowing a static description of the cage. With the decrease in temperature, the correlation between structure and dynamics increases. The study shows that at lower temperatures, the structural heterogeneity increases, and since the structure becomes a better predictor of the dynamics, it leads to an increase in the dynamical heterogeneity. We also find that the softness of a hard, immobile region evolves with time and becomes soft and eventually mobile due to the rearrangements in the neighbourhood, confirming the well-known facilitation effect.

\end{abstract}

\maketitle
\section{Introduction}
When a liquid is cooled fast, it evades crystallization, and below the onset temperature enters a supercooled liquid domain where the properties of the liquid appear quite different from that at high temperatures. Although the average structure across the onset temperature appears to change continuously, the dynamics shows a marked difference and below the onset temperature, it slows down substantially and also becomes heterogeneous \cite{ediger_spatial_heterogenity, kob_plimpton_dynamic_heterogenity}.
The origin of this dynamic heterogeneity in supercooled liquids is a topic of intense research \cite{ediger_spatial_heterogenity, kob_plimpton_dynamic_heterogenity,harowell_nature_phy,harrowell_prl_2006,cooper_free_volume,liu_nature,rottler,cubuk}. In analogy with crystals where under external perturbation regions with structural defects show a higher probability of rearrangements \cite{taylor_plastic_deformation, collings_colloid_crystal}, it is often suggested that in supercooled liquids such structural defects may also be present, giving rise to the dynamic heterogeneity.  However, while identifying structural defects in crystals with otherwise well-defined structure is trivial, doing the same in a supercooled liquid where particles are arranged in a disordered manner is a non-trivial task \cite{jp_dyre_structure}. 

Harowell and coworkers investigated different properties of the initial configuration of a supercooled liquid and analyzed their correlation with irreversible rearrangements in the system \cite{harowell_nature_phy,harrowell_prl_2006,cooper_free_volume}. They found that the local inherent structure potential energy \cite{harrowell_prl_2006} and free volume\cite{cooper_free_volume} does not have any correlation with these rearrangements, but the Debye-Waller Factor\cite{harrowell_prl_2006} and the low-frequency normal modes of the systems are spatially correlated with the irreversible structural rearrangements\cite{harowell_nature_phy}. Liu and Manning have shown that under shear, the rearrangements of particles take place in the regions which contribute to these low-frequency normal modes\cite{manning_vib_modes}.  
Smessaert and Rottler did a quantitative analysis of these low-frequency soft modes and showed that they are long-lived  \cite{rottler}. There are also studies where the dynamics is connected to the elastic properties of the system \cite{shoving,local_elasticity_jean_louis,lerne,zaccone_pnas}. The heterogeneity of the local elastic modulus was found to be correlated with the dynamic heterogeneity showing regions with low shear modulus have higher plastic activity \cite{local_elasticity_jean_louis}. A recent study has proposed that the slowing down of the dynamics is controlled by the mesoscopic elastic stiffness parameter, which is more sensitive than the shear modulus \cite{lerne}.

Apart from the above mentioned correlation between vibrational and elastic modes with the dynamics, there are also other studies that calculate properties that are purely structural in nature and correlate them with the dynamics. Using mutual information technique, it was shown that coarse-grained energy and density correlates with the dynamics \cite{paddy}. Another technique that has been quite successful in the recent time is Machine learning (ML). ML techniques allow for the identification of a softness parameter that encodes in a single number a large number of structural descriptors \cite{cubuk,liu_nature,olivier_PRE}.  It was shown that below the onset temperature, the dynamics is controlled by this softness parameter\cite{liu_nature} and this softness parameter for attractive and repulsive systems are different, which eventually leads to the
difference in their dynamics \cite{olivier_PRE}. It was also shown that this softness parameter could
identify structural defects which are similar to dislocation in the crystalline solids, and ductile materials have a large number of soft spots compared to brittle materials \cite{richard_prm}. A recent experimental study of colloidal glasses showed that this ML softness parameter can predict the
devitrification process \cite{ganpati_NP_2021}. Unfortunately, this softness consists of a linear reweighting of the local pair correlation, the weights being blindly found by the ML algorithm and its physical connection to structure remains unclear. 

Interestingly in the liquid state theories \cite{hansen} like mode coupling theory \cite{gotze_condmat} and density functional theory \cite{wolynes_kirkpatrick} the structure plays an important role in determining the dynamics. However, in the supercooled liquid regime, since the change in the structure is small and gradual, whereas the dynamics slows down dramatically, the validity of these theories has been questioned. Also, studies showing that in this regime, systems with similar structures have orders of magnitude differences in their dynamics further supports the idea that in this regime structure does not play a role in the dynamics \cite{berthier_tarjus_prl_2009,bertheir_tarjus_EPJE_2011}. However certain extensions of these theories have been found to work reasonably well in the supercooled regime \cite{sarika_PNAS,chong_pre,role_pair_configuration,manoj_PRL,indranil,Schweizer_2003_jcp,schweizer_2005_jcp,mei_pnas_2021,ghosh_schweizer_jcp2020,mirigian_schweizer_jcp2014,mei_schweizer}. In a study involving some of us the dynamic density functional theory (DDFT) \cite{wolynes_kirkpatrick,schweizer_2005_jcp} was used to develop a microscopic mean field theory \cite{role_pair_correlation}. It was shown that the softness of the mean field caging potential described by the structure of the liquid is connected to the dynamics for a wide variety of systems, even for those which are out of equilibrium \cite{manoj_PRL}. However all these theories have their root in the liquid state theory and thus only deal with the average properties of the system. Also note that both softness and dynamics can have a local microscopic variation and it is well known that the correlation between average quantities does not guarantee a correlation at a microscopic level. Thus using liquid state theories to study local properties and a causal relationship between structure and dynamics at the local level is a nontrivial and challenging task.

In this paper, we study the correlation between softness and dynamics at a microscopic level. Using DDFT formalism we first describe the softness parameter at a local level in terms of the structure of the liquid. We then show that this microscopic softness can capture the structural heterogeneity, and the lifetime of the softness parameter is similar to the lifetime of the cage. This, we believe, is a nontrivial. The most important result is the observed causal relationship between local softness and local rearrangements. 

The rest of the paper is organized as follows. In the next section, we provide the simulation details. In Sec. III, we present the formalism we use to identify local rearrangements. Sec. IV presents the details of the calculation of the softness parameter at the local level and looks at its distribution and time evolution.  In Sec. V we study the correlation between the local rearrangement events and the local softness. Finally, in Sec. VI, we present the conclusion.

\section{Simulation Details}

The system we studied is the Kob-Andersen model for glass-forming liquid, which is a binary mixture (80:20) of Lennard-Jones (LJ) particles \cite{PRL_73_1376_1994}. The interaction between the particles i and j, where i,j=A,B (the type of the particles) is given by

\begin{equation}
 U_{ij}(r)=
\begin{cases}
 U_{ij}^{(LJ)}(r;\sigma_{ij},\epsilon_{ij})- U_{ij}^{(LJ)}(r^{(c)}_{ij};\sigma_{ij},\epsilon_{ij}),    & r\leq r^{(c)}_{ij}\\
   0,                                                                                       & r> r^{(c)}_{ij},
\end{cases}
\label{pot}
\end{equation}
\noindent
where $u(r_{ij})$ = $4\epsilon_{ij}[(\sigma_{ij}/r_{ij})^{12}-(\sigma_{ij}/r_{ij})^6]$, $r_{ij}$ is the distance between particles i and j and $\sigma_{ij} $ is effective diameter of particle and $r_{i,j}^{(c)}=2.5\sigma_{ij}$. The length, temperature, and time are given in units of $\sigma_{AA}$ , $\epsilon_{AA}/k_B$, $(m\sigma_{AA}^2 /\epsilon_{AA})^{1/2}$, respectively.

We use $\sigma_{AA}=1.0 $, $\sigma_{AB}=0.8 $, $\sigma_{BB}=0.88 $, $\epsilon_{AA}=1.0 $, $\epsilon_{AB}=1.5 $, $\epsilon_{BB}=0.5 $, $ m_{A}=m_{B}=1 $ and Boltzmann constant  $k_{B} = 1$.
In our simulations we have used periodic boundary conditions and Nos\'{e}-Hoover thermostat  with integration timestep 0.0025$\tau$. The time constants for  Nos\'{e}-Hoover thermostat  are taken to be 100  timesteps. The total number density $\rho=N/V=1.2$ is fixed, where V is the system volume, and N=4000 is the total number of particles.

\section{Identifying Rearrangements}
In this work, we aim to correlate structure parameter with the mobility at a local level; for identifying fast particle or an event, there are many ways which can be used, such as doing iso-configurational runs and identify irreversible reorganisations\cite{harowell_nature_phy} or tracking the mean square displacement over a period of time. Here we have used a method which was first proposed by Candelier et al.\cite{candelier,PRE_88_022314_2013} where they calculate a quantity $p_{hop}(i,t)$ which captures for every particle i in a certain time window W = $[t_{1},t_{2}]$, the cage jumps when the average position of the particle changes rapidly. The expression for $p_{hop}(i,t)$ is    
\begin{equation}
\begin{aligned}
 p_{hop} & (i,t) = \\
 & \sqrt{<(\vec{r_{i}} - <\vec{r_{i}}>_{U})^{2}>_{V}^{1/2}<(\vec{r_{i}} - <\vec{r_{i}}>_{V})^{2}>_{U}^{1/2}} ,
\end{aligned}
\end{equation}
for all t $\in$ W, where averages are performed over the time intervals surrounding time t, i.e., U = [$t-\Delta t/2$, t] and V = [t, $t+\Delta t/2$] where $\Delta t$ should be a timescale over which the particles can rearrange.
For a time window W the small value of $p_{hop}$ means the particle is contained within same cage and conversely if $p_{hop}$ is large this means particle is within two distinct cage (see APPENDIX A for details). To identify a rearrangement event we set a threshold value $p_{c}$, as done in Ref.\cite{olivier_PRE} where $p_{c}$ is chosen as the root mean squared displacement $<\Delta r(t)^{2}>$ value, computed at a time where the non-Gaussian parameter \scalebox{0.95}{$\alpha_{2}$ = $\frac{3<\Delta r^{4}(t)>}{5<\Delta r^{2}(t)>^{2}} - 1 $} has a maximum. When $p_{hop}>p_{c}$ we consider that a rearrangement event has taken place.
  The $p_{c}$ values are 0.115 for T=0.47, 0.130 for T=0.53, 0.141 for T=0.58, 0.159 for T=0.7, and 0.178 for T=0.8.  We also vary $\Delta t$ from $15$ to $75$ LJ units and find that qualitatively the results remain quite similar. For the rest of the work we consider $\Delta t=15$ LJ unit.
 We also find that the neighbourhood of the fast particles are structurally less ordered (see APPENDIX B for details).

\section{Computing Local Softness}

We present a brief sketch of the DDFT formalism which is discussed in detail in earlier studies \cite{role_pair_correlation,manoj_PRL,schweizer_2005_jcp}. The time evolution of the density, under mean-field approximation, can be written in terms of a Smoluchowski equation in an effective mean-field caging potential \cite{role_pair_correlation,wolynes_kirkpatrick, schweizer_2005_jcp} which is obtained from the Ramakrishnan-Yussouff free energy functional \cite{ry_form}. 

\begin{figure}
\centering
\begin{subfigure}{0.33\textwidth}
\includegraphics[width=0.9\linewidth]{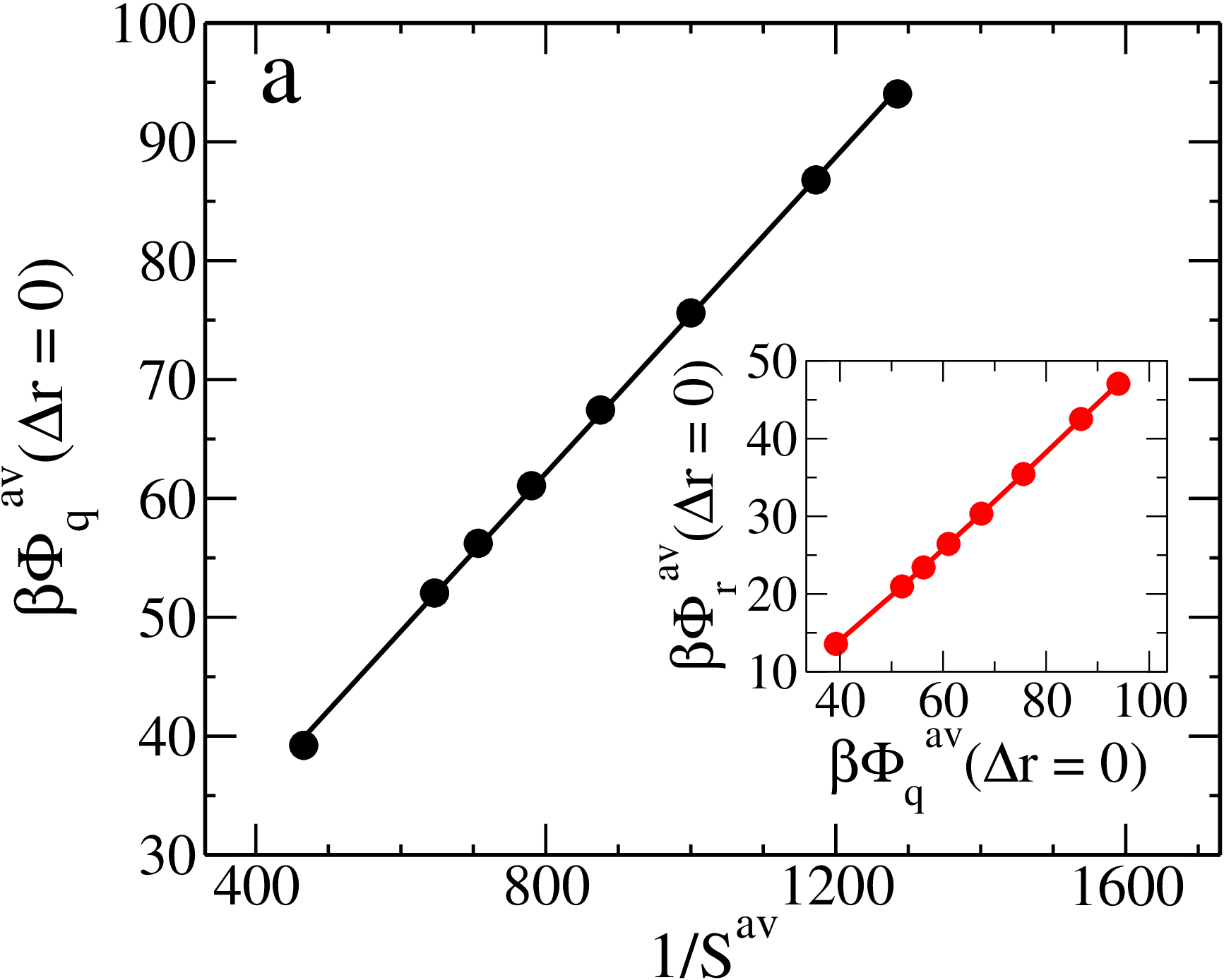}
%,trim = {0 0cm 0 0.1cm},clip]{final_phir_phiq.eps}
\end{subfigure}
\vskip\baselineskip
\hspace*{0.4cm}
\begin{subfigure}{0.33\textwidth}
\includegraphics[width=0.9\linewidth]{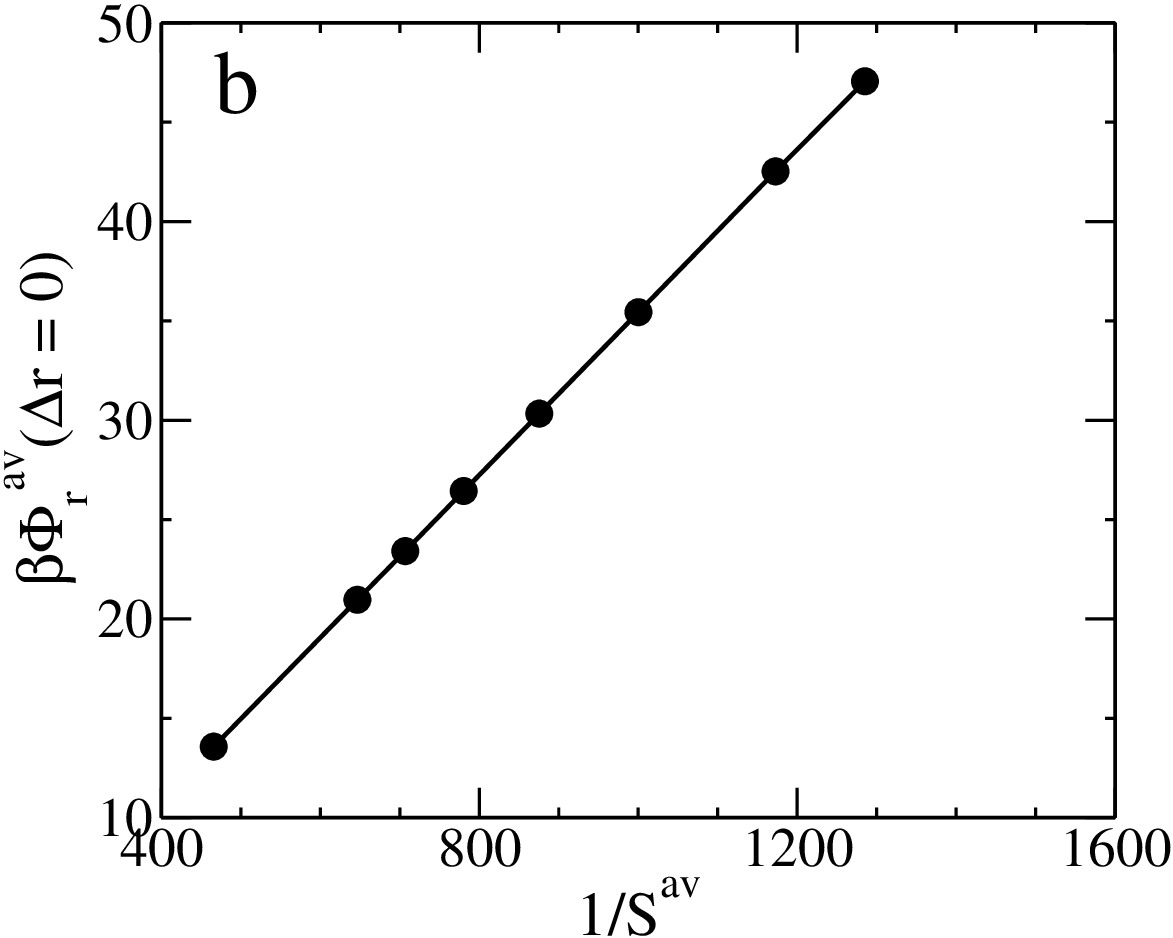}
%,trim = {0 0cm 0 0.1cm},clip]{final_1byS_phiq.eps}
\end{subfigure}
\caption{(a) $\beta \Phi_{q}^{av}(\Delta r=0)$ is plotted against $1/S^{av}$ at different temperatures and in inset $\beta \Phi_{r}^{av}(\Delta r=0)$ is plotted against $\beta \Phi_{q}^{av}(\Delta r=0)$ at different temperatures, 
they both show linear proportionality. (b) $\beta \Phi_{r}^{av}(\Delta r=0)$ $vs$ $1/S^{av}$ is also linear.}
\label{Fig_linear}
\end{figure}

The mean-field potential is written as,

\begin{equation}
\begin{aligned}
\beta & \Phi^{av}_{q}(\Delta r) = \\ 
& -\int\frac{{\bf{dq}}}{(2\pi)^{3}}\sum_{uv}C_{uv}(q)\sqrt{x_{u}x_{v}}[S_{uv}(q) - \delta_{uv}]e^{\frac{-q^{2}{\Delta r}^{2}}{6}} . 
\end{aligned}
\label{old_softness}
\end{equation}
\noindent
This formalism is similar to that used by Schweizer and coworkers\cite{Schweizer_2003_jcp,schweizer_2005_jcp,schweizer_2003_trans_coeff}. Here $\Delta r$ is the displacement of the central particle from its equilibrium position. $\beta=1/k_BT$ and $x_{u/v}$ represent the fraction of particle of type A/B in the binary mixture. In the above expression, $S_{uv}({\bf{q}})$ = $(1/\sqrt{N_{u}N_{v}})\sum_{i=1}^{N_{u}}\sum_{j=1}^{N_{v}}$ $exp[-i{\bf{q.}}({\bf{r}}_{i}^{u}$-${\bf{r}}_{j}^{v})$ and ${\bf{C(q) = 1 - S^{-1}(q)}}$, where ${\bf{S(q)}}$ is the partial structure factor of the liquid and ${\bf{C(q)}}$ and ${\bf{S(q)}}$ are in matrix form. Note that this is a mean-field potential, and the assumption is that the cage described by the structure of the liquid remains static when the particle moves by a distance $\Delta r$. The superscript $'av'$ implies that these are global quantities averaged over all particles and also over ensembles.  
  To quantify the average softness of the potential, we fit the potential near $\Delta r = 0$ to a harmonic form, $\beta\Phi^{av}(r)$ = $\beta\Phi_{q}^{av}(\Delta r = 0)+\frac{1}{2}(\Delta r)^{2}/S^{av}$, where $S^{av}$ is the softness parameter and $\Phi_{q}^{av}(\Delta r = 0)$ is the value of the potential at the minima,
\begin{equation}
\begin{split}
 \beta\Phi_{q}^{av}&( \Delta r=0)  = \\
 & -\int{\frac{{\bf{dq}}}{(2\pi)^{3}}\sum_{uv}C_{uv}(q)\sqrt{x_{u}x_{v}}(S_{uv}(q) - \delta_{uv})} .  
\end{split}
\label{depth_q}
\end{equation}
\noindent
The subscript $q$ in $\Phi_{q}^{av}(\Delta r=0)$ implies that for the calculation of the depth of the potential, the parameters are first expressed in the wavenumber plane and then integrated over `q'. From our earlier study\cite{manoj_PRL} and, also as shown in Fig.\ref{Fig_linear}(a) we find that the softness is inversely proportional to the depth of the potential $\Phi_{q}^{av}(\Delta r=0)$. We can write $1/S^{av}=a_{0}+a_{1}\Phi_{q}^{av}(\Delta r=0)$ where the parameters $a_{0}$ and $a_{1}$ appear to be constant.
  
In this present work, the aim is to study the softness at the local level, {\it i.e.} the softness of each particle in a single time frame. According to our theory, this will require calculating the structure factor in a single snapshot for each particle. The calculation of structure factor without time and particle averaging leads to a sharply fluctuating non-integrable function. Thus we cannot use Eq.\ref{depth_q} to calculate the depth of the mean caging potential at the local level.
  
However, this equation can also be written as an integration over the radial distance and is given by,
\begin{equation}
\begin{aligned}
\beta\Phi_{r}^{av}(\Delta r=0) = -\rho\int{{\bf{dr}}\sum_{uv}C_{uv}(r)\sqrt{x_{u}x_{v}}g_{uv}(r)} ,
\end{aligned}
\label{depth_r}
\end{equation}
\noindent
where $\rho$ is the density, $g_{uv}(r)$ is the partial radial distribution function and $C_{uv}(r)$ is the partial direct correlation function in real space.
Here again, the subscript `r' in $\beta\Phi_{r}^{av}(\Delta r=0)$ implies that for the calculation of the depth of the potential, the functions are first expressed in the `r' plane and then integrated over `r'. This `r' denotes the distance between the central tagged particle whose softness is being calculated and its neighbours. We should not confuse it with $\Delta r$ in Eq.\ref{old_softness}, which denotes the displacement of the tagged particle within its cage.

Interestingly, although Eq.\ref{depth_q} and Eq.\ref{depth_r} are derived from a microscopic theory, we can arrive at the same expression from an intuitive argument. The direct correlation function represents the short-range effective interaction potential between two particles \cite{hansen}. Thus the product of the direct correlation function and the structure information should provide the effective two-body level caging potential of the tagged particle.

The depth of the potential can be expressed both in terms of the structure factor and the radial distribution function (rdf) (Eq.\ref{depth_q} and Eq.\ref{depth_r}).  In Eq.\ref{depth_r} apart from the rdf we need the information of the direct correlation function. This is usually calculated by Fourier transforming $C(q)$, which in turn depends on $S(q)$. Since the aim here is to formulate a theory that can calculate the local softness and, as discussed before, calculating $S(q)$ at a local level is not possible; thus, for the calculation of $C(r)$, we make an approximation. In the integral equation theory, the hypernetted chain (HNC) approximation \cite{hansen} provides an expression for $C(r)$ in terms of the rdf and the interaction potential,
\begin{equation}
\begin{aligned}
C_{uv}(r) = -\beta U_{uv}(r) + (g_{uv}(r) - 1) -ln(g_{uv}(r)) ,
\end{aligned}
\label{direct}
\end{equation}
\noindent
where $U_{uv}(r)$ is the interaction potential given by Eq.\ref{pot}, which is an input to the theory. Here we neglect the bridge function present in the HNC approximation. Using this approximate form for $C(r)$ we calculate $\Phi_{r}^{av}(\Delta r=0)$ and compare it with $\Phi_{q}^{av}(\Delta r=0)$ (inset Fig.\ref{Fig_linear}(a)). Since we use an approximate form for $C(r)$ we find that $\Phi_{r}^{av}(\Delta r=0) \neq \Phi_{q}^{av}(\Delta r=0)$ but they are proportional to each other. This can be expressed as $\Phi_{q}^{av}(\Delta r=0)= b_{0} +b_{1}\Phi_{r}^{av}(\Delta r=0)$. Note that if we know the values of $b_{0}$ and $b_{1}$  which appear to be independent of temperature, then from the approximate value of the depth of the potential, $\Phi_{r}^{av}(\Delta r=0)$, we can obtain the actual value of the depth of the potential, $\Phi_{q}^{av}(\Delta r=0)$.  We can then exploit the relation between $\Phi_{q}^{av}(\Delta r=0)$ and $S$ to obtain the softness parameter. Thus we can write,
\begin{eqnarray}
\begin{aligned}
1/S^{av}=&a_{0}+a_{1}*(b_{0}+b_{1}\Phi_{r}^{av}(\Delta r=0))\\
=&c_{0}+c_{1}\Phi_{r}^{av}(\Delta r=0),
\end{aligned}
\label{soft}
\end{eqnarray}
\noindent
where $c_{0}=a_{0}+a_{1}*b_{0}$ and $c_{1}=a_{1}*b_{1}$. 
This predicted linearity between the softness and the approximate depth of the potential $\Phi_{r}^{av}(\Delta r=0)$ is shown to be valid in Fig.\ref{Fig_linear}(b).  
The values of the parameters are independent of temperature, and we obtain them by calculating the depth of the potential and its softness at the whole system level. We now assume that Eq.\ref{soft}  also describe the relationship between the local softness and the depth of the local potential with the same values of the parameters ($c_{0}$ and $c_{1}$). Thus to calculate the softness at a local level, we need to calculate the local $\Phi_{r}(\Delta r=0)$. The parameters without the superscript $'av'$ describes them at a local level. Note that the expression of the depth of the local caging potential is same as Eq.\ref{depth_r}, but the rdf and the direct correlation function, which are input in the calculation, are now obtained at the local level.

\noindent

Although at local level, we cannot calculate $S(q)$, we can calculate the rdf. The single-particle rdf in single frame can be expressed as a sum of Gaussians and is given by \cite{piaggi}.
\begin{equation}
\begin{aligned}
 g_{\mu\nu}^{i}(r) = \frac{1}{4\pi \rho r^{2}} \sum_{j}\frac{1}{\sqrt{2\pi\delta^{2}}}e^{-\frac{(r - r_{ij})^{2}}{2\delta^{2}}},
\end{aligned}
\label{rdf1}
\end{equation}
\noindent
where $\delta$ is the variance of Gaussian distribution. The variance is used to make the otherwise discontinuous function a continuous one. In this calculation we assume $\delta=0.09\sigma_{AA}$. The smaller the value of $\delta$, the more accurate is the description of the rdf and thus the potential (see APPENDIX B for details).

Using Eqs.\ref{depth_r}, \ref{direct} and \ref{rdf1} we calculate the $\Phi_{r}(\Delta r=0)$ at a local level. Then using Eq.\ref{soft} we calculate the softness at a local level where $c_{0}$ and $c_{1}$ are predetermined from the correlation of $S^{av}$ and $\Phi^{av}_{r}(\Delta r=0)$. The results shown here are only for the `A' particles. (further calculation details of $\Phi_{r}(\Delta r=0)$ is given in the APPENDIX B).

\begin{figure}
\centering
\begin{subfigure}{0.33\textwidth}
\includegraphics[width=0.9\linewidth]{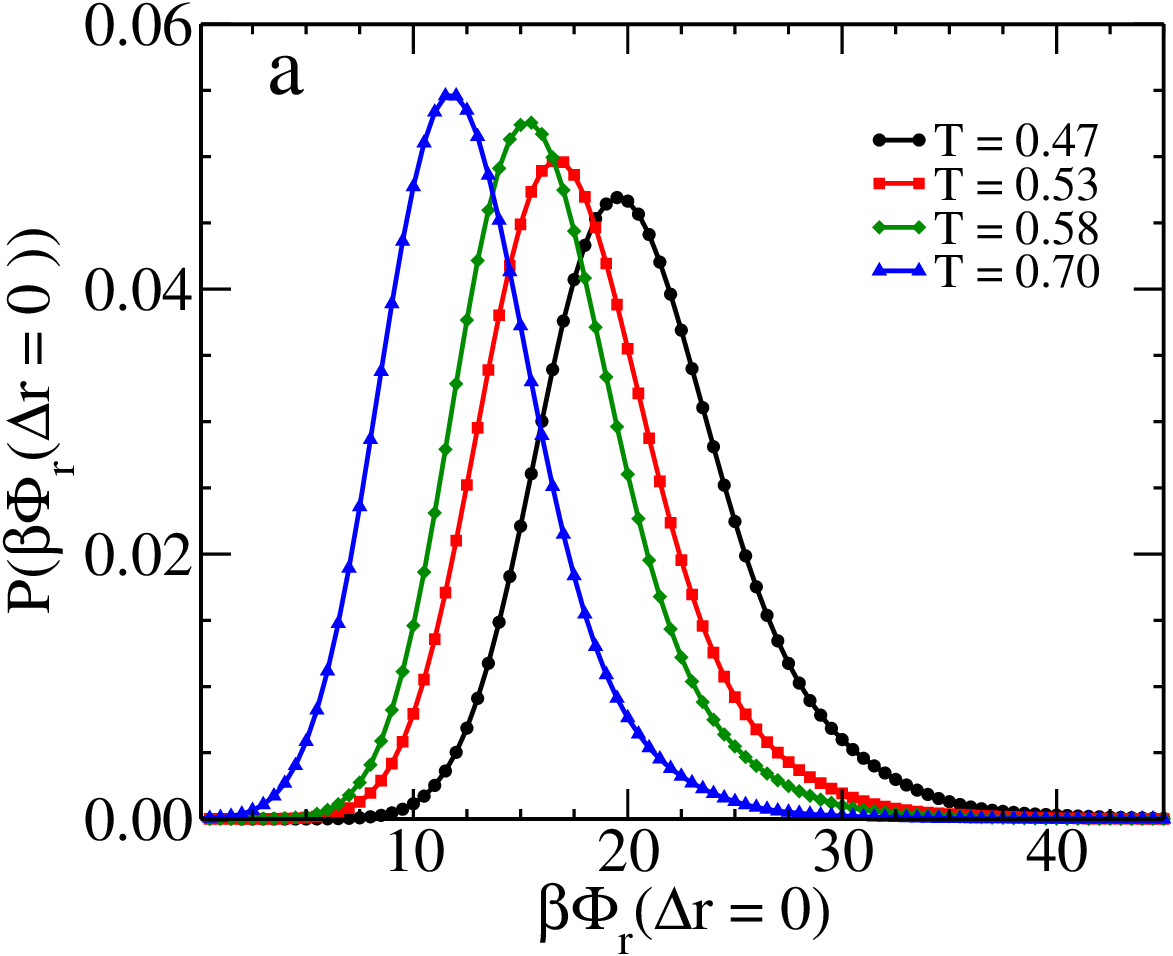}
\end{subfigure}
\vskip\baselineskip
\begin{subfigure}{0.33\textwidth}
\includegraphics[width=0.9\linewidth]{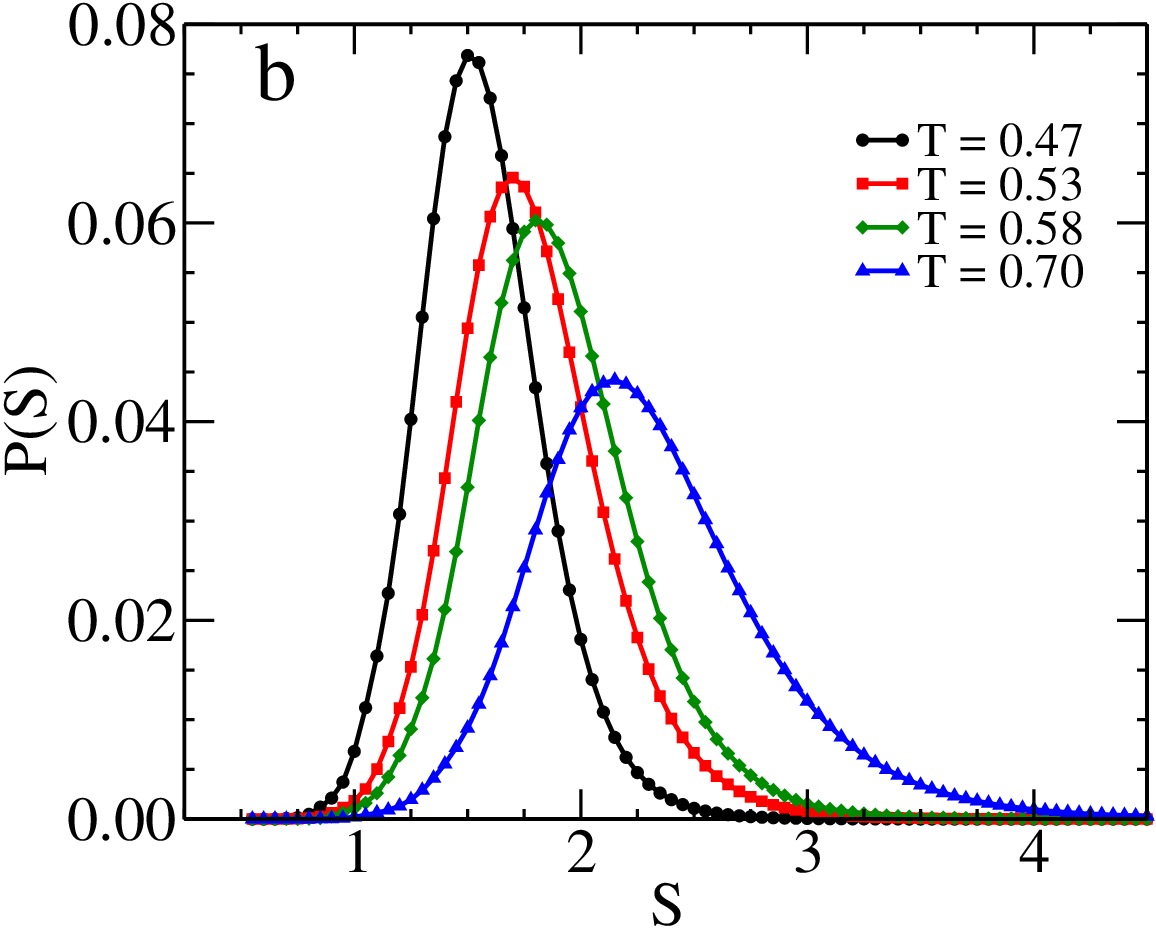}
\end{subfigure}
\caption{(a) Distribution of $\beta \Phi_{r}(\Delta r=0)$ at different temperatures. $\Phi_{r}(\Delta r=0)$ is calculated from Eq.\ref{depth_r} using the local $g(r)$ (see APPENDIX B) . (b) Distribution of softness for different temperatures. The softness values mentioned in the axis properties are scaled by $10^{-3}$.}
\label{distribution}
\end{figure}

\subsection{Distribution of local softness}
In Fig.\ref{distribution}(a) we plot the distribution of $\beta\Phi_{r}(\Delta r=0)$.
With decrease in temperature the distribution  broadens and shifts to higher values. Thus our formalism captures the heterogeneity in the structure and its temperature variation. The broadening of the distribution at lower temperatures predicts that with a decrease in the temperature, there is an increase in the heterogeneity of the structure. The shift to higher values is an effect of the structure becoming more well defined at lower temperatures leading to a deeper caging potential. 
\begin{figure*}[t]
\centering
\begin{subfigure}{0.32\textwidth}
\includegraphics[width=0.9\linewidth]{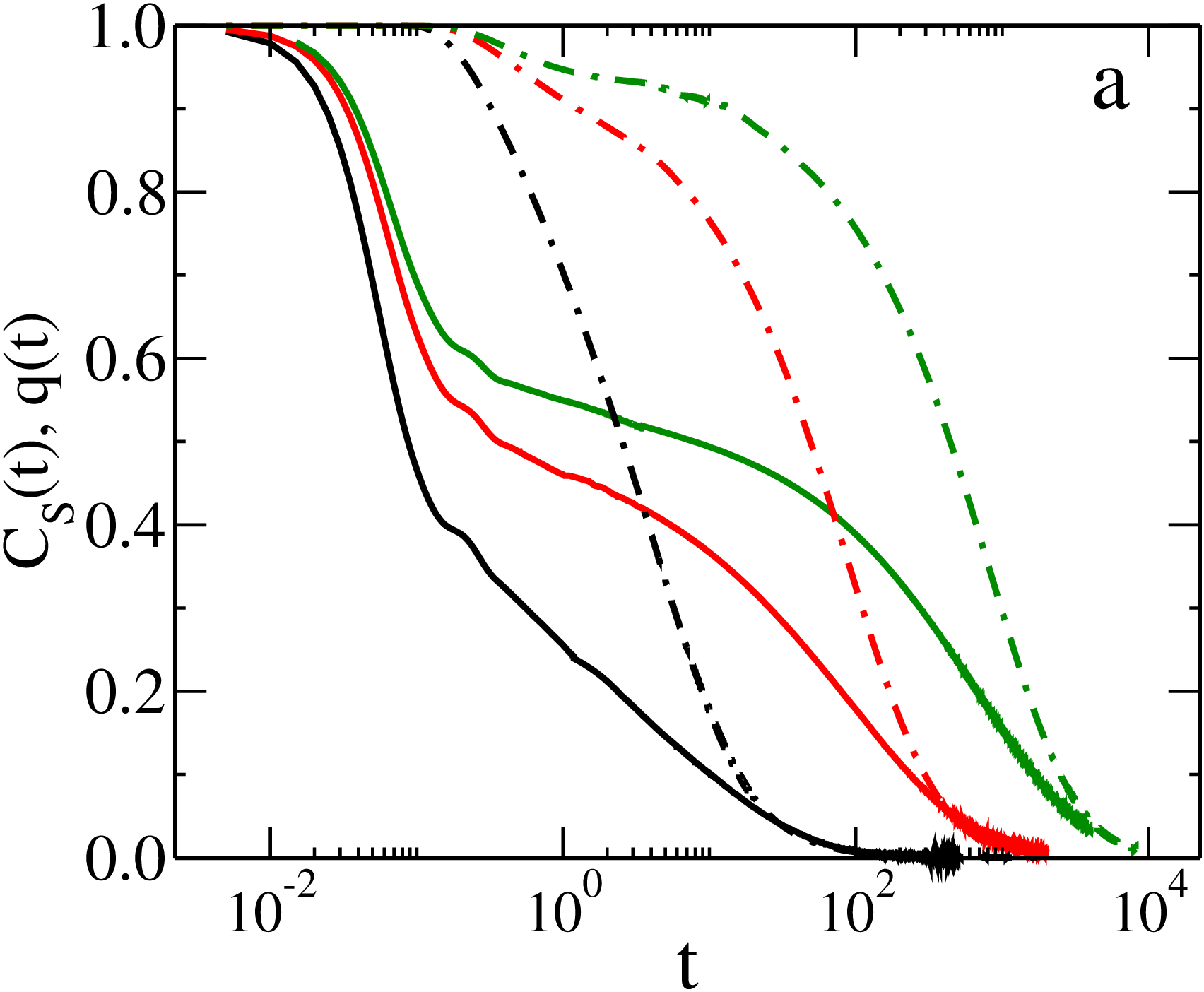}
\end{subfigure}
%\hskip\baselineskip
\begin{subfigure}{0.32\textwidth}
\includegraphics[width=0.9\linewidth]{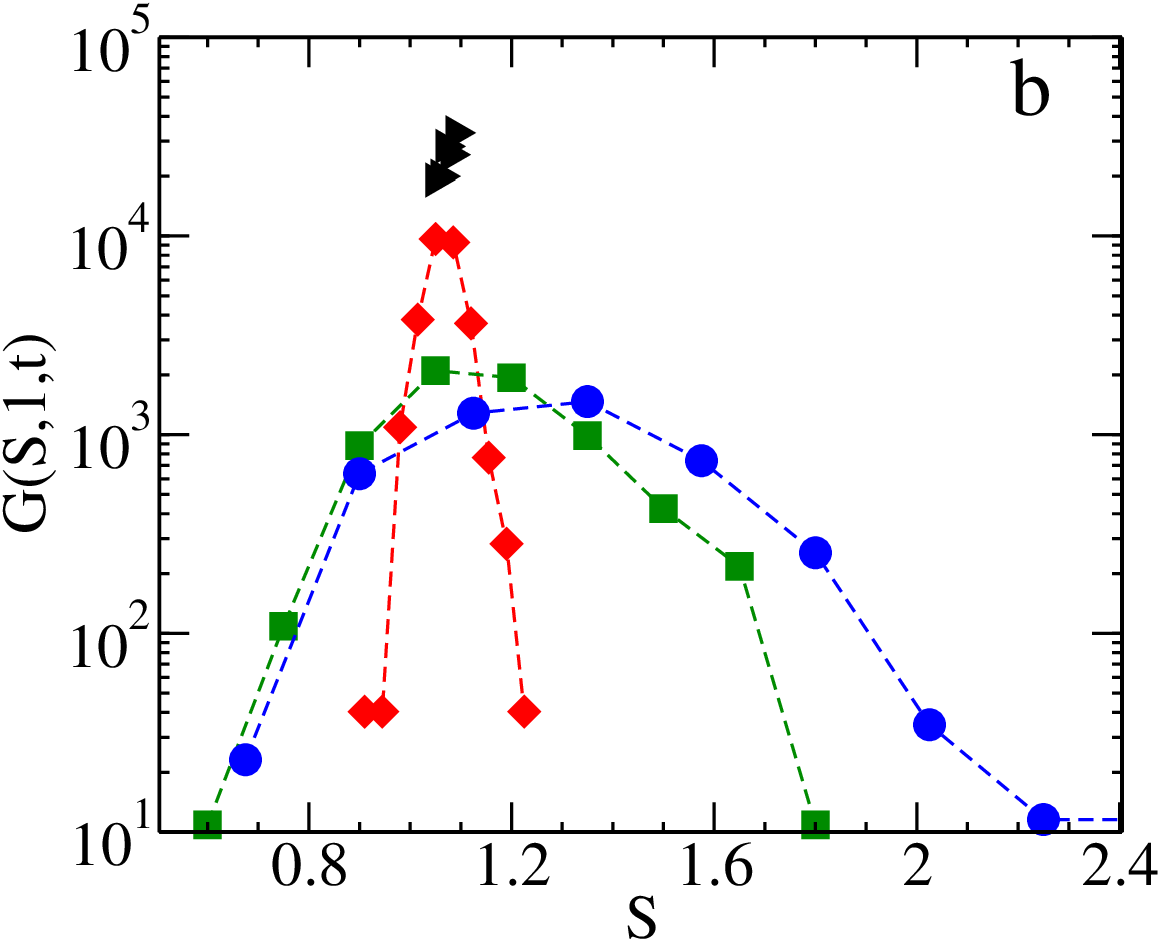}
\end{subfigure}
\begin{subfigure}{0.31\textwidth}
\includegraphics[width=0.9\linewidth]{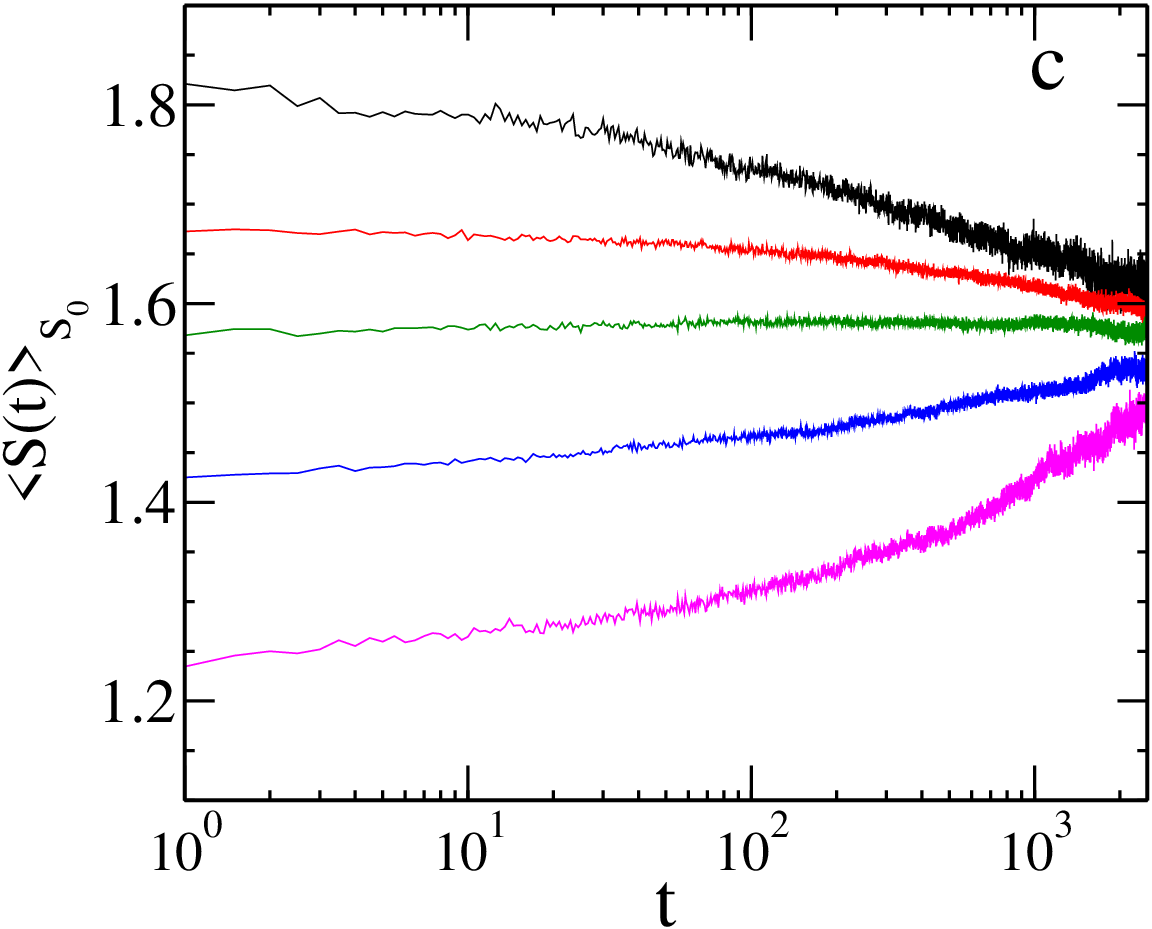}
\end{subfigure}
%\begin{figure*}
%\centering
%\begin{subfigure}{0.32\textwidth}
%\includegraphics[width=0.9\linewidth]{fig2a.eps}
%\end{subfigure}
%%%\vskip\baselineskip
%\begin{subfigure}{0.35\textwidth}
%\includegraphics[width=0.9\linewidth]{scaled_pre_fig4b.eps}
%\end{subfigure}
%\begin{subfigure}{0.31\textwidth}
%\includegraphics[width=1\linewidth]{scaled_pre_fig4c.eps}
%\end{subfigure}
\caption{\textbf (a) Time evolution of softness for different temperatures. For comparison we also plot the overlap function. The solid lines are $C_{S}(t)$ and dotted lines are q(t) values. The temperatures studied are 0.80(black), 0.53(red) and 0.47(dark-green). (b) The time evolution of the softness propagator, $G(S,S_{0},t)$ for a collection of particles that do not move beyond $0.5$ till time `t'. The initial softness $S_{0}$ $\approx$ 0.001. The data is obtained at t = 0(black), t = 0.01(red), t= 10(dark-green) and t = 1000 (blue). (c) The time evolution of the average softness for particles that do not move beyond $0.5$ till time `t'. This is plotted for several groups of particles which have initial softness values ranging from $S_{0} \approx $ 0.002 (black) to $S_{0} \approx $ 0.001.  (magenta). The softness values mentioned in the axis properties are scaled by $10^{-3}$.}
\label{propogator}
\end{figure*}

Exploiting the relationship between the depth of the caging potential and the softness we obtain the distribution of softness $P(S)$ at different temperatures (Fig.\ref{distribution}(b)). The study shows that with lowering of temperature, the distribution moves to lower values of softness and also becomes narrower. The narrowing of the $P(S)$ at lower temperatures might appear to contradict the increase in structural heterogeneity. However, note that this is a combined effect of the inverse correlation between the depth of the mean-field potential and the softness and the fact that at lower temperatures  $P(\beta\Phi_{r}(\Delta r=0))$ shifts to higher values of $\beta\Phi_{r}(\Delta r=0)$. This narrowing of softness distribution at lower temperatures can also be seen in the ML studies \cite{olivier_PRE}.   

The shift of the softness distribution with temperature is similar to but more prominent than that observed in the ML study \cite{liu_nature} and different from the study where the softness was expressed as low-frequency vibrational modes, and the distribution was found to be almost temperature-independent\cite{rottler}. To the best of our knowledge, such a strong temperature effect of the softness distribution has not been seen in other existing formalism. However, in one of the earlier studies, such shift in the local (per molecule) inherent structure energy distribution with temperature was observed \cite{sciortino}.

\subsection{Time evolution of local softness}

For the softness field to have some effect on the dynamics, it has to have a finite lifetime.  Following an earlier work \cite{rottler} we define the time evolution of the softness field as the average of the Pearson correlation given by,  $C_{S}(t) = \Big <\frac{[S_{i}(t')- \bar{S}(t') ][S_{i}(t+t')- \bar{S}(t+t')]}{\sigma_{S(t')}\sigma_{S(t+t')}} \Big >$ 
 where $\bar{S}(t)=1/N \sum_{i} S_{i}(t)$ is average softness value at time $t$ over all particles and $\sigma_{S(t)}$ is the standard deviation of the softness at time t. The final average given by the angular brackets is over time origin, $t'$, and also over particles.
%\begin{figure}[h]
%\centering
%\subfigure{\includegraphics[width=.35\textwidth]{final_overlap_vs_softnessdecay_difftemp.eps}}
%\subfigure{\includegraphics[width=.35\textwidth]{latest_4b.eps}}
%\subfigure{\includegraphics[width=.35\textwidth]{4c.eps}}
%\caption{}
%\label{crosspoint_PDI}
%\end{figure}

In Fig.\ref{propogator}(a) we plot the time evolution of the softness, $C_{S}(t)$, and compare it with the overlap function, $q(t)$ given by, $q(t) =\frac{1}{N} \Big \langle \sum_{i=1}^{N}  \omega (|{\bf{r}}_i(t)-{\bf{r}}_i(0)|)\Big \rangle \quad $
where function $\omega(x)$ is 1 when $0\leq x\leq a$ and $\omega(x)=0$ otherwise. The cut-off parameter $a=0.3$ is chosen such that particle positions separated
due to small amplitude vibrational motion are treated as the
same and $a^2$ is comparable to the value of the MSD in the plateau between the ballistic and diffusive regimes \cite{kobandersenLJ}. The $\alpha$ relaxation time $\tau_{\alpha}$ is defined such that $q(t=\tau_{\alpha})=1/e$. We find that the time evolution of the softness shows a similar two-step decay as the overlap function. The plateau becomes more prominent at lower temperatures. This connects the softness parameter to the local cage around a particle. Ideally, we would expect that when a particle moves leading to the decay of the overlap function, the local softness around that particle changes. However, the softness field can change even when the particle does not move out of the cage. This change in the softness field around a particle happens due to rearrangements in the neighbourhood.  Following the ML study\cite{liu_nature}, we define the softness propagator $G(S,S_{0},t)$ which describes the time evolution of the softness of particles that move less than a distance `0.5' till time `t'. In Fig.\ref{propogator}(b) we plot $G(S,S_{0},t)$ at $T=0.47$ where we choose particles whose initial value of softness $S_{0}=0.001$. Note that in terms of the softness parameter value, these are hard (low softness) particles. We see that with progress in time, the softness distribution of these particles, which is initially sharply peaked because of our choice of particles, becomes wider, and the peak position shifts to the right. Eventually, the distribution reaches the average form. This evolution of softness of particles which are confined in a region of `$\Delta r=0.5$' leads to the initial sharp drop in $C_{S}(t)$. 

In Fig.\ref{propogator}(c) we plot the time evolution of the average softness, $<S(t)>_{S_{0}}$ for those particles which start with the softness value $S_{0}$ and displace less than the distance `0.5' at a time `t'. We find that the softness evolves with time and at times larger than the alpha relaxation time ($\tau_{\alpha}=700$ at T=0.47) all of them reach the average value. This evolution of the softness field has also been observed in the ML study \cite{liu_nature}.

%\begin{figure}[!hbt]
%\centering
%\vspace*{-20mm}
%\hspace*{-10mm}
%\includegraphics[width=0.35\textwidth,trim = %{0 0cm 0 0.1cm},clip]{different_temp_combined%_time_evol.eps}
%\caption{Time evolution of softness for %different temperatures} 
%\label{time}
%\end{figure}

\section{Correlation between local rearrangement and local softness}
\begin{figure}
\centering
\begin{subfigure}{0.33\textwidth}
\includegraphics[width=0.9\linewidth]{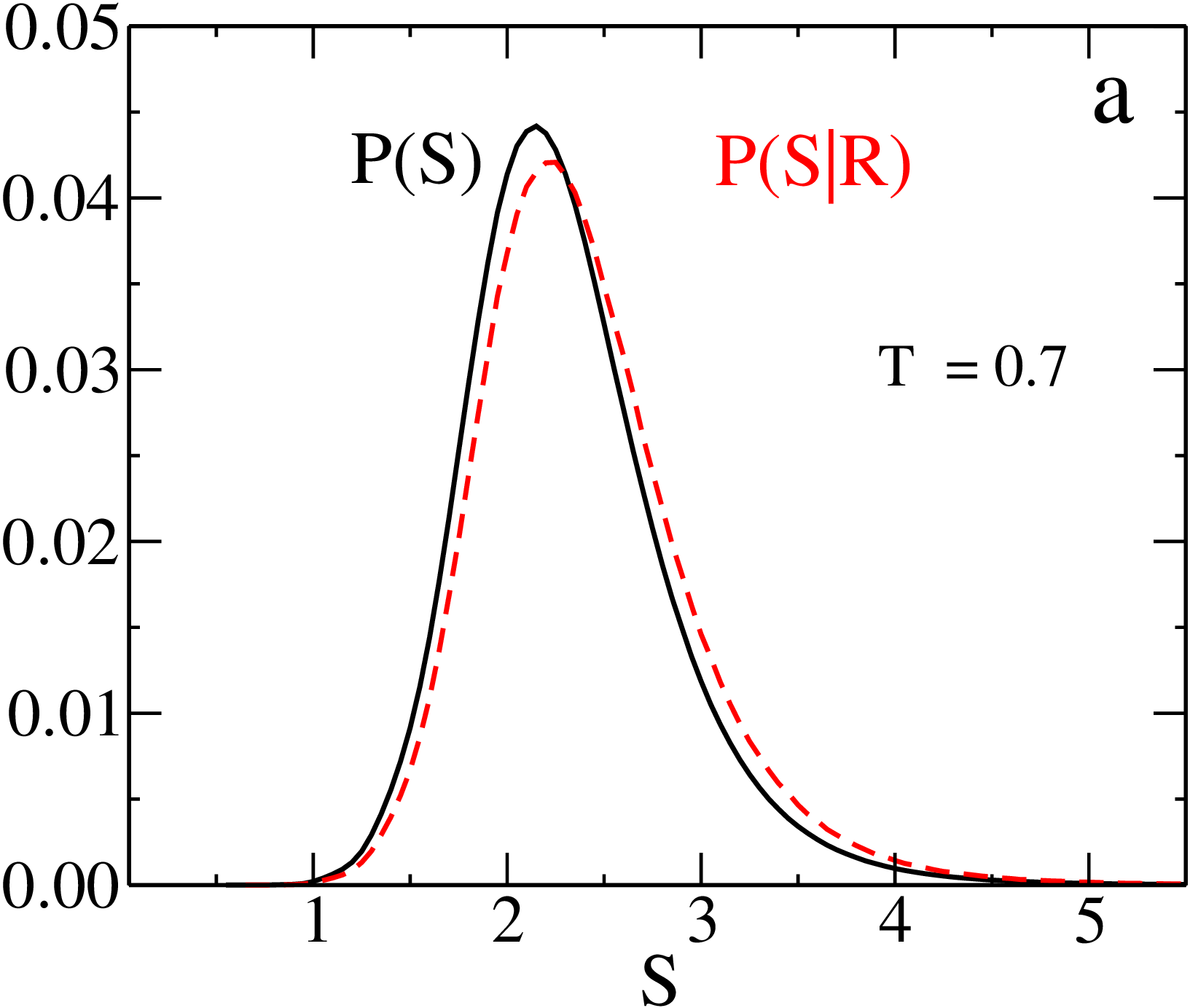}
\end{subfigure}
\vskip\baselineskip
\begin{subfigure}{0.33\textwidth}
\includegraphics[width=0.9\linewidth]{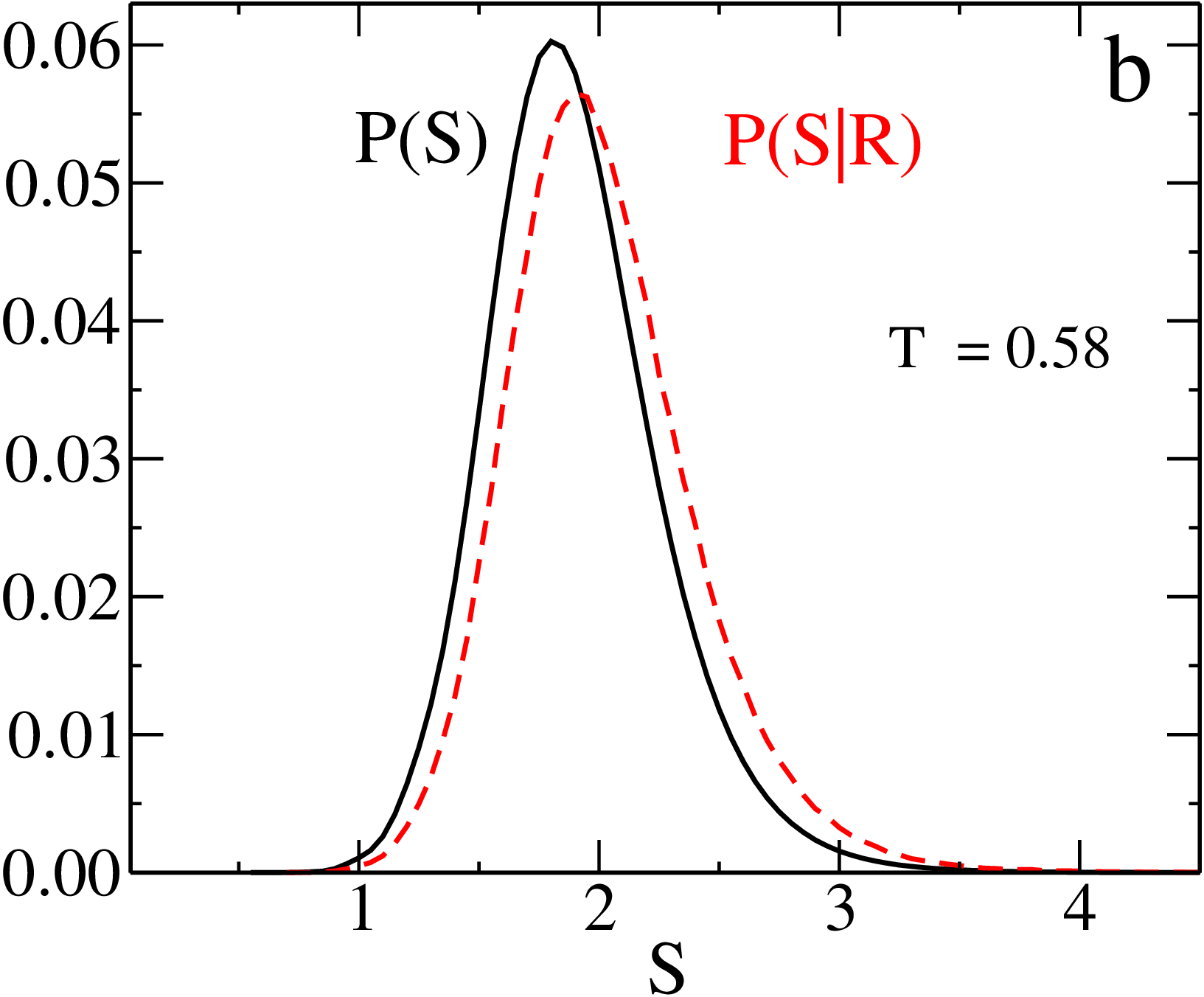}
\end{subfigure}
\vskip\baselineskip
\begin{subfigure}{0.33\textwidth}
\includegraphics[width=0.9\linewidth]{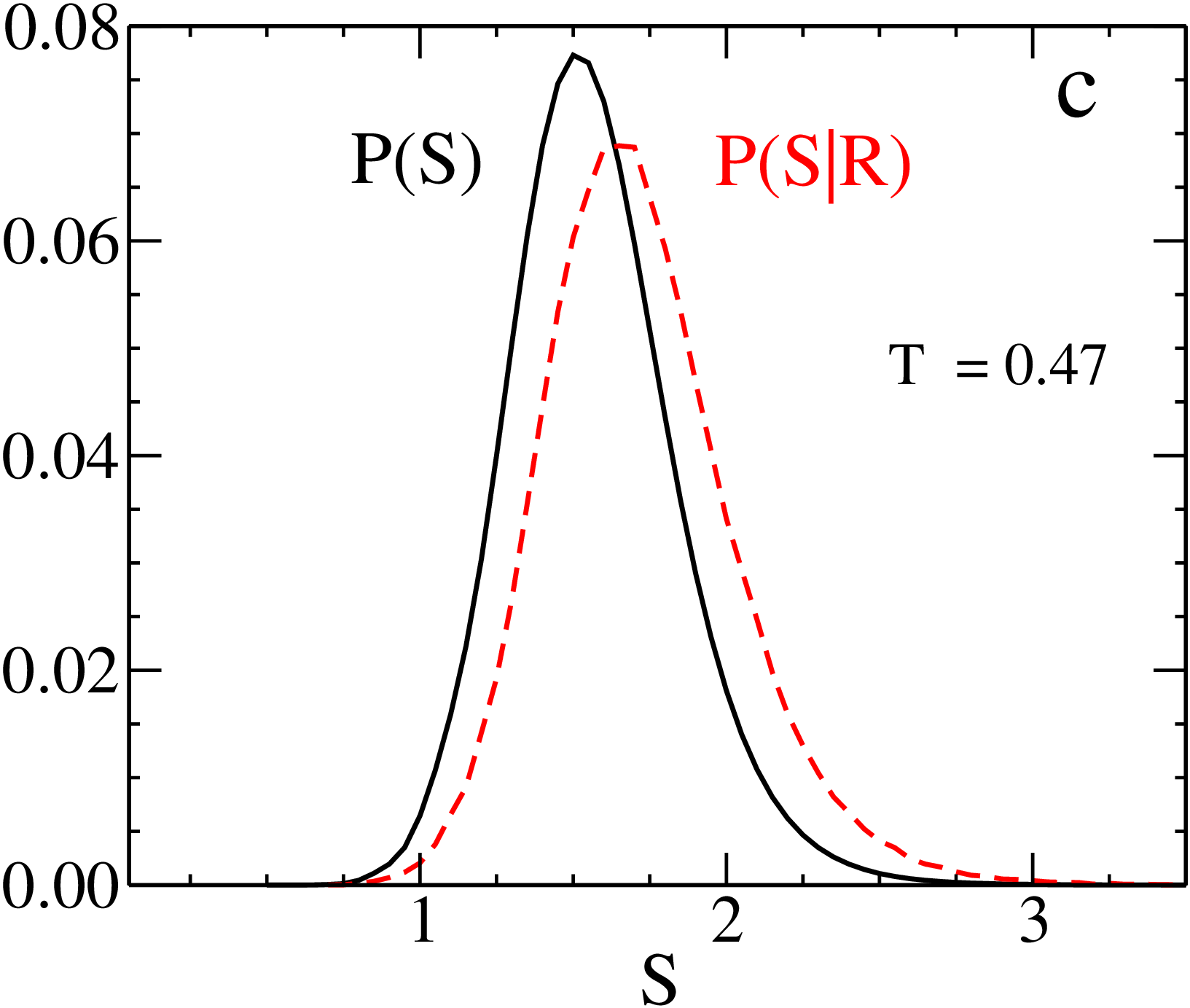}
%,trim = {0 0cm 0 0.1cm},clip]{final_1byS_phiq.eps}
\end{subfigure}
\caption{ The distribution of softness of all particles, $P(S)$ in the system(black) and of those which are about to rearrange(red) at different temperatures. The softness values mentioned in the axis properties are scaled by $10^{-3}$.}
\label{fast_phop}
\end{figure}

The main objective of this work is to identify a parameter that can describe structural heterogeneity and understand its role in dynamical heterogeneity. We have already shown that the softness of the mean-field potential can explain structural heterogeneity. We also find that the lifetime of this softness parameter is comparable with the timescale of the system dynamics. We now put this parameter through a more stringent test and study the causal relationship between the softness and dynamics. The dynamics expressed via local rearrangement events is computed using a well-established method \cite{olivier_PRE,candelier,candelier_prl,PRE_88_022314_2013}. We find that particles that undergo local rearrangements have a less structured neighbourhood (see APPENDIX B for details). We also find that
the percentage of particles
with a softness value greater than the peak, which undergoes rearrangement increases with a decrease in temperature. At T=0.7 it is 65\%, at T=0.58 it is 69\% and at  T=0.47 it is 75\%. (see Fig.\ref{fast_phop}). 

\begin{figure*}
\centering
\begin{subfigure}{.36\textwidth}
\includegraphics[width=0.9\linewidth]{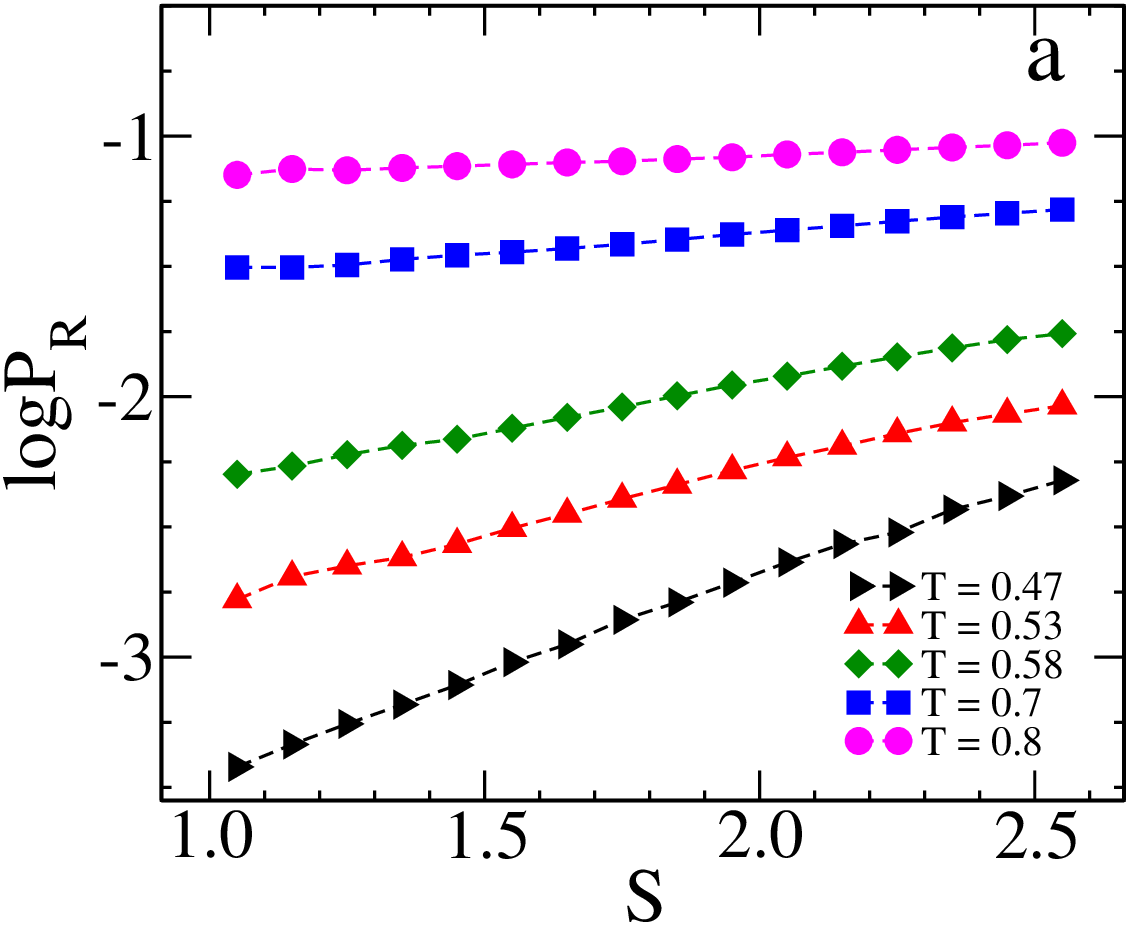}
\end{subfigure}
\begin{subfigure}{.36\textwidth}
\includegraphics[width=0.9\linewidth]{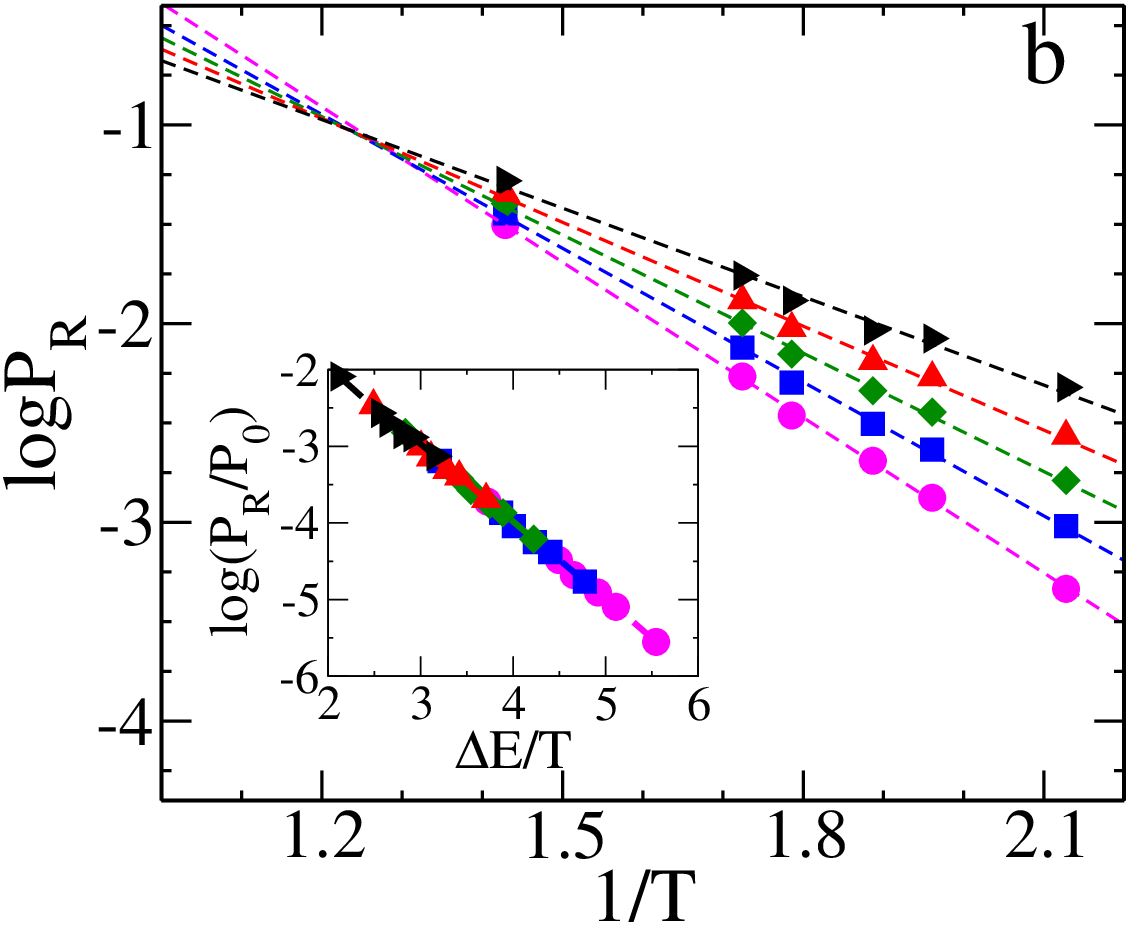}
\end{subfigure}
%\vskip\baselineskip
\vskip\baselineskip
\begin{subfigure}{.36\textwidth}
%%\hspace*{0.5}
\includegraphics[width=0.9\linewidth]{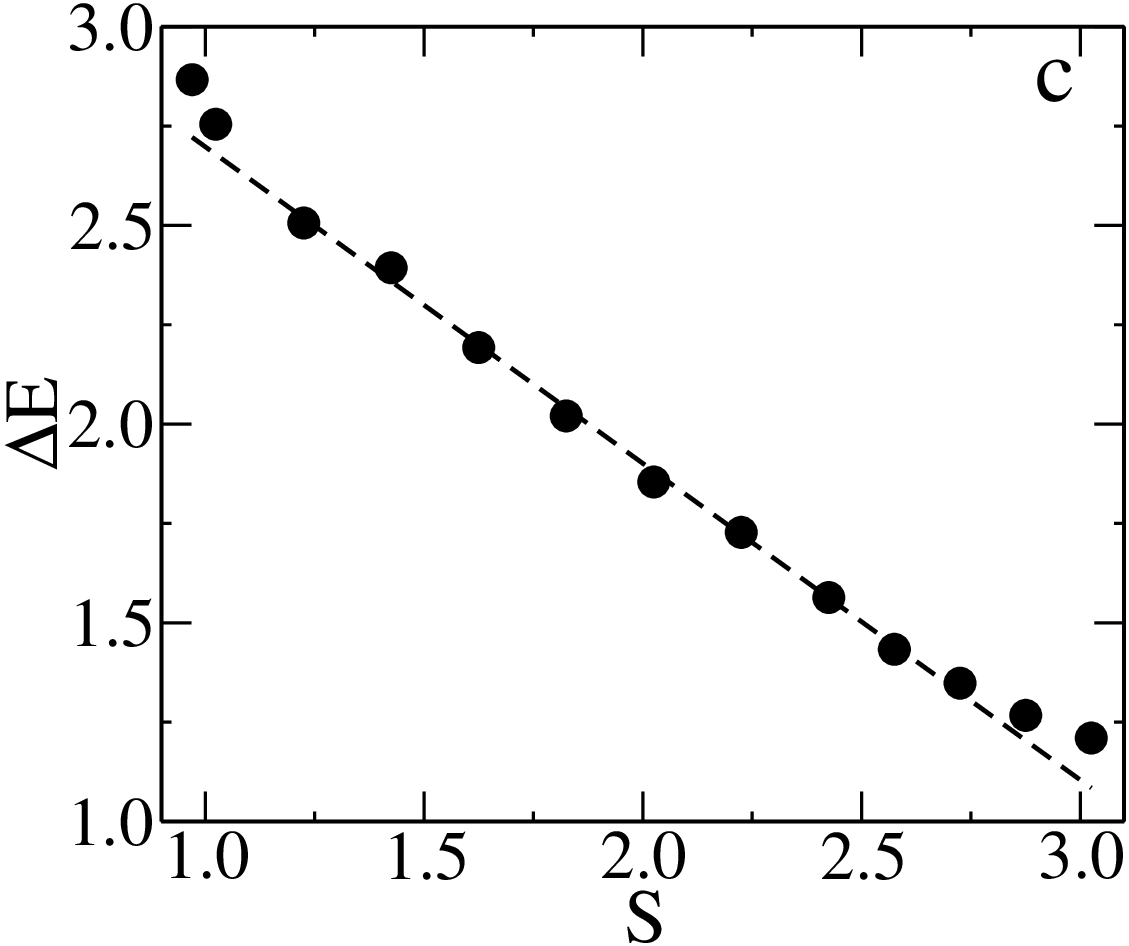}
\end{subfigure}
\begin{subfigure}{.38\textwidth}
\includegraphics[width=0.9\linewidth]{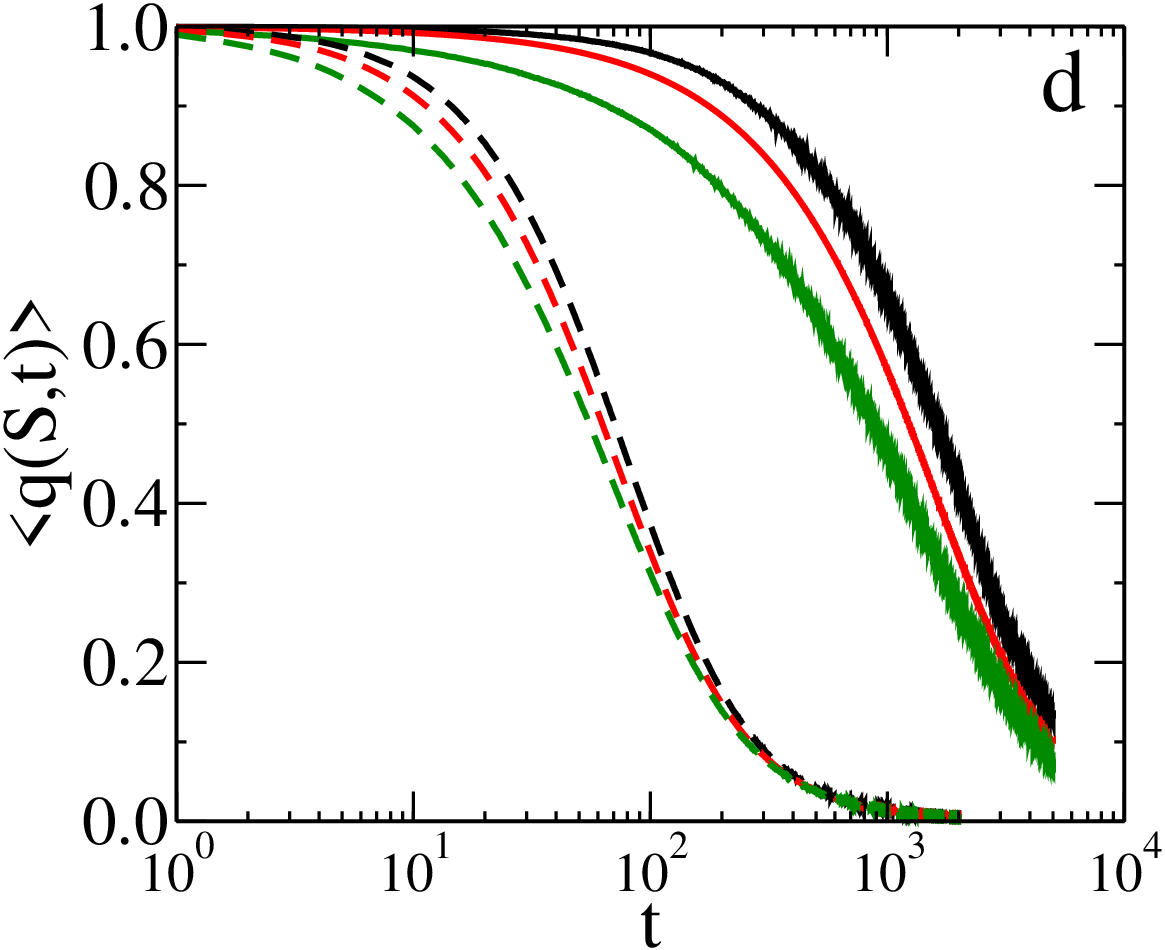}
\end{subfigure}
\caption{\textbf  (a) The softness dependence of the fraction of particles with softness value $S$ that undergo rearrangement, $P_{R}(S)$, for different temperatures. (b) $P_{R}(S)$ as a function of 1/T for five different softness values from S $\approx$ 0.001 (magenta) to S $\approx$ 0.0025 (black). $\Delta E$ is obtained from this plot assuming $P_{R}(S)$ = $exp(- \Delta E/T)$. The inset shows the collapse of these probabilities when $P_{R}/P_{0}$ is plotted against $\Delta E/T$.  (c)  The softness dependence of energy barrier, $\Delta E$. (d)$q(S,t)$ at different softness values (highest,lowest and moderate), and at two temperatures T=0.58(dashed line) and T=0.47(solid line). The softness values mentioned in the axis properties are scaled by $10^{-3}$. The
base of log is 10.} 
\label{pr_vs_S_T}
\end{figure*}

To quantify the correlation between softness value and a rearrangement event, we calculate the fraction of particles having a specific value of softness that undergoes rearrangement, $P_{R}(S)$ as a function of $S$ at different temperatures (Fig.\ref{pr_vs_S_T}(a)). We find that at all temperatures, $P_{R}(S)$ has a softness dependence. Particles with higher values of softness have a higher rearranging probability.

This dependence becomes orders of magnitude stronger at lower temperatures. This shows that at lower temperatures, the dynamics get more coupled to the local structure. We have already demonstrated that with a decrease in temperature, the structural heterogeneity increases. We believe that these two factors, i) increase in structural heterogeneity ii) larger coupling between structure and dynamics, together lead to the well-known increase in the dynamic heterogeneity at lower temperatures. Particles with small softness values (confined in a deeper mean-field potential) are less mobile compared to particles that have larger values of softness (confined in a shallower mean-field potential). 
 We now check if the probability of a local rearrangement can be expressed in an Arrhenius form $\it i.e.$ $P_{R}(S)=P_{0}(S) exp(-\Delta E(S)/T)$ where the activation energy is a function of softness. $log_{10} P_{R}(S)$ as a function of inverse temperature for different softness values show a linear behaviour (Fig.\ref{pr_vs_S_T}(b)), thus confirming the Arrhenius form. The energy barrier, $\Delta E(S)$, the slopes of the lines are a function of $S$. $P_{R}/P_{0}$ against $\Delta E(S)/T$ (inset of Fig.\ref{pr_vs_S_T}(b)), shows a data collapse, thus further validating the Arrhenius form. At smaller values of softness, $P_{R}(S)$ strongly depends on $T$. In Fig.\ref{pr_vs_S_T}(b), we find that the extrapolated lines cross each other at a certain temperature, and above this temperature, it appears that $P_{R}(S)$ increases with a decrease in $S$, which is an unphysical result. Thus above this temperature, where the extrapolated lines cross the correlation between $P_{R}(S)$ and softness does not remain valid. 
Interestingly the extrapolated lines cross at $T=0.801$ which is similar to the onset temperature of the system \cite{atreyee_onset}. Below we argue why it should be related to the onset of the glassy dynamics.

The soft spots are equivalent to the defects in the crystals. Thus they should come from a solid-like description of the system.
When describing the softness, we assume a well-defined cage that remains static over the time period that we are calculating the dynamics. We have also shown that the lifetime of the softness has a correlation with the lifetime of the cages. It is well-known that in supercooled Kob-Andersen model liquid, these cages appear below the onset temperature, where the decoupling between the short and long time dynamics starts \cite{sastry_debenedetti,kob_mctheory1995}. Thus our theoretical formalism rightly predicts that the correlation between softness and dynamics starts around the onset temperature, and at lower temperatures where there is a larger decoupling between the short and long time dynamics leading to a longer lifetime of the cages, the local softness becomes a better predictor of the local rearrangement events. 

 In Fig.\ref{pr_vs_S_T}(c) we plot the activation energy, $\Delta E(S)$ as a function of S, and it shows that as softness decreases, the activation energy increases. The range of energy in our study (1.2-2.9) is narrower than that obtained in the ML study (5.0-11.0)\cite{liu_nature}, but the energy values are similar. There are two possible reasons why our study predicts a narrower range. i) To calculate the activation energy for different softness values, we need to plot a temperature dependence of $P_{R}(S)$ which can be done in the range of softness where the  $P(S)$ at different temperatures overlap. Since $P(S)$ in our study shows a larger shift with temperature, this overlapping range is narrower than the ML study. ii) Our calculation of softness is obtained from a mean-field microscopic theory,  thus bound by the two-body microscopic correlation functions. It is possible that beyond what the theory predicts, other correlation functions contribute to the softness parameter and are picked up in the weight function of the ML study. However, as discussed in the ML study, our study reveals that the dominant contribution comes from the two-body terms. In Fig.\ref{pr_vs_S_T}(d), we plot the softness dependent overlap function, $q(S,t)$ at different softness values (highest, lowest, and moderate) and at two temperatures. The time scale of the overlap function shows a stronger softness dependence at lower temperatures. This clearly shows a causal relationship between the softness and the dynamics that manifests itself more at lower temperatures.

\section{Conclusion}

In this work, we describe the softness at the local level. We find that the softness parameter predicts the presence of structural heterogeneity, which grows at lower temperatures and is also longer lived. The lifetime of the softness parameter is correlated with the well-known cage structure in the supercooled liquids. We establish a causal relationship between the local softness and the local rearrangement events. We further show that even for particles that do not undergo rearrangement, the softness parameter evolves in time due to the dynamics in the neighbourhood, giving rise to the well-known facilitation effect \cite{chandler_pnas2003,chandler_2010annurev}.
An immobile hard region can eventually become soft and have a higher probability of becoming mobile due to rearrangements in the neighbourhood.  

The results presented here are strikingly similar to that obtained in the ML work, although the methodology of obtaining the softness in the two different studies are entirely different \cite{liu_nature}. As suggested by the Authors themselves in the ML study, since the softness depends on a large number of local parameters, it is difficult to interpret the physical meaning of the softness \cite{liu_improved_softness}. The conjecture is that through the multiple local parameters, the information of the local cage around particles is incorporated in the softness; however, there is no direct proof of it. The advantage of the present study is it calculates the softness from a microscopic theory that describes the caging potential in terms of the structure of the liquid. The depth of the caging potential is similar to the local energy that is obtained by reweighting the local pair correlation by the direct correlation function. We find that there exists a causal relationship even between the depth of this local caging potential and the dynamics (see APPENDIX B for details). Given the direct correlation of the present softness parameter and the liquid structure, it will be interesting to understand its correlation with other existing definitions of the soft fields like the softness parameter obtained in machine learning studies, the quasi localized vibrational modes and the elastic stiffness parameter. These problems will be addressed in future studies. \\

{\bf Acknowledgement}\\

S.~M.~B thanks Chandan Dasgupta, Olivier Dauchot and Jeppe Dyre for discussion, and acknowledges funding support from Science and Engineering Research Board (SERB), Department of Science and Technology Government of India. M.~S. thanks Ujjwal Kumar Nandi, and Palak Patel for the help with the initial setup of the system and discussions and thanks SERB for fellowship.\\ %\\[4mm]

\textbf{APPENDIX A: Identifying rearrangements}\\

In this work, we aim to correlate structure parameter with the mobility at a local level; for identifying fast particle or an event, there are many ways which can be used, such as doing iso-configurational runs and identify irreversible reorganisations\cite{harowell_nature_phy} or tracking the mean square displacement over a period of time. Here we have used a method which was first proposed by Candelier et al.\cite{candelier,PRE_88_022314_2013} where they calculate a quantity $p_{hop}(i,t)$ which captures for every particle i in a certain time window W = $[t_{1},t_{2}]$, the cage jumps when the average position of the particle changes rapidly. The expression for $p_{hop}(i,t)$ is    

\begin{equation}
\begin{aligned}
 p_{hop} & (i,t) = \\
 & \sqrt{<(\vec{r_{i}} - <\vec{r_{i}}>_{U})^{2}>_{V}^{1/2}<(\vec{r_{i}} - <\vec{r_{i}}>_{V})^{2}>_{U}^{1/2}} ,
\end{aligned}
\end{equation}
for all t $\in$ W, where averages are performed over the time intervals surrounding time t, i.e., U = [$t-\Delta t/2$, t] and V = [t, $t+\Delta t/2$] where $\Delta t$ should be a timescale over which the particles can rearrange.
\begin{figure}[!hbt]
\centering
\includegraphics[width=0.30\textwidth,trim = {0 0cm 0 0.1cm},clip]{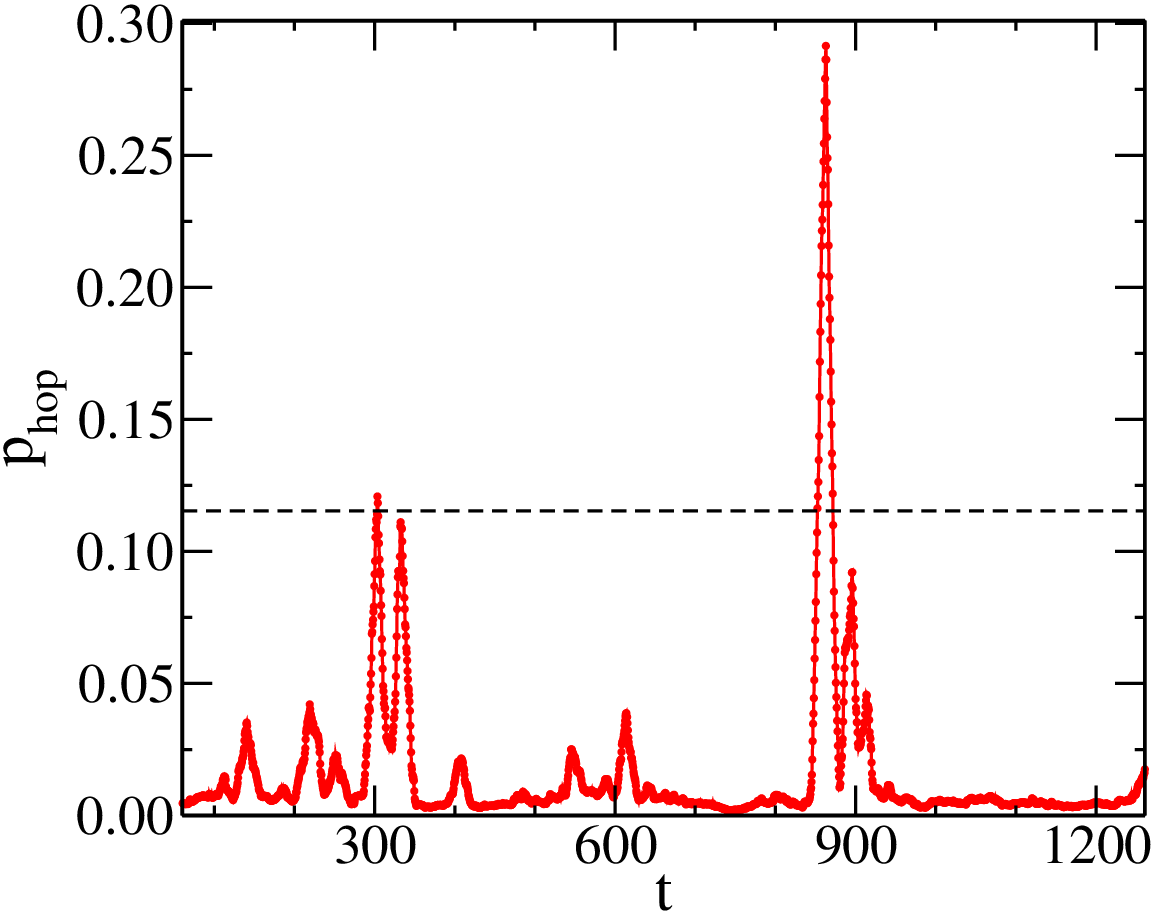}
\caption{The value of $p_{hop}$ as a function of time for a representative particle at T=0.47. The dashed black line is the $p_{c}$ value at T = 0.47. When $p_{hop}>p_{c}$ we consider a rearrangement event has taken place.}
\label{phop}
\end{figure}
 For a time window W the small value of $p_{hop}$ means the particle is contained within same cage and conversely if $p_{hop}$ is large this means particle is within two distinct cage. To identify a rearrangement event we set a threshold value $p_{c}$, as done in Ref.\cite{olivier_PRE} where $p_{c}$ is chosen as the root mean squared displacement $<\Delta r(t)^{2}>$ value, computed at a time where the non-Gaussian parameter \scalebox{0.95}{$\alpha_{2}$ = $\frac{3<\Delta r^{4}(t)>}{5<\Delta r^{2}(t)>^{2}} - 1 $} has a maximum. When $p_{hop}>p_{c}$ we consider that a rearrangement event has taken place.
 The $p_{c}$ values are 0.115 or T=0.47, 0.130 for T=0.53, 0.141 for T=0.58, 0.159 for T=0.7, and 0.178 for T=0.8.  We also vary $\Delta t$ from $15$ to $75$ LJ units and find that qualitatively the results remain quite similar. For the rest of the work we consider $\Delta t=15$ LJ unit.
 In Fig.\ref{phop} we plot the $p_{hop}$ for a particle which shows few rearrangement events.

\begin{figure}
\centering
\begin{subfigure}{0.23\textwidth}
\includegraphics[width=0.9\linewidth]{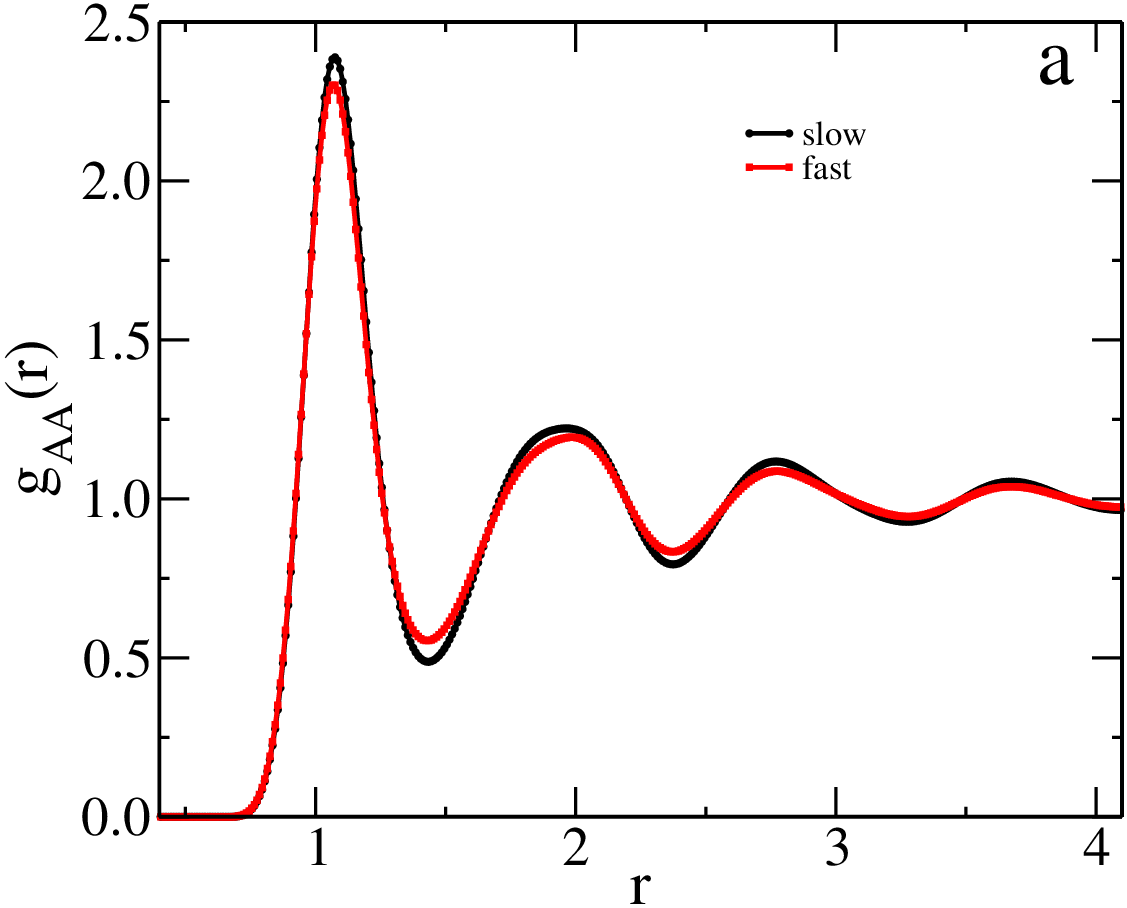}
\end{subfigure}
\begin{subfigure}{0.23\textwidth}
\includegraphics[width=0.9\linewidth]{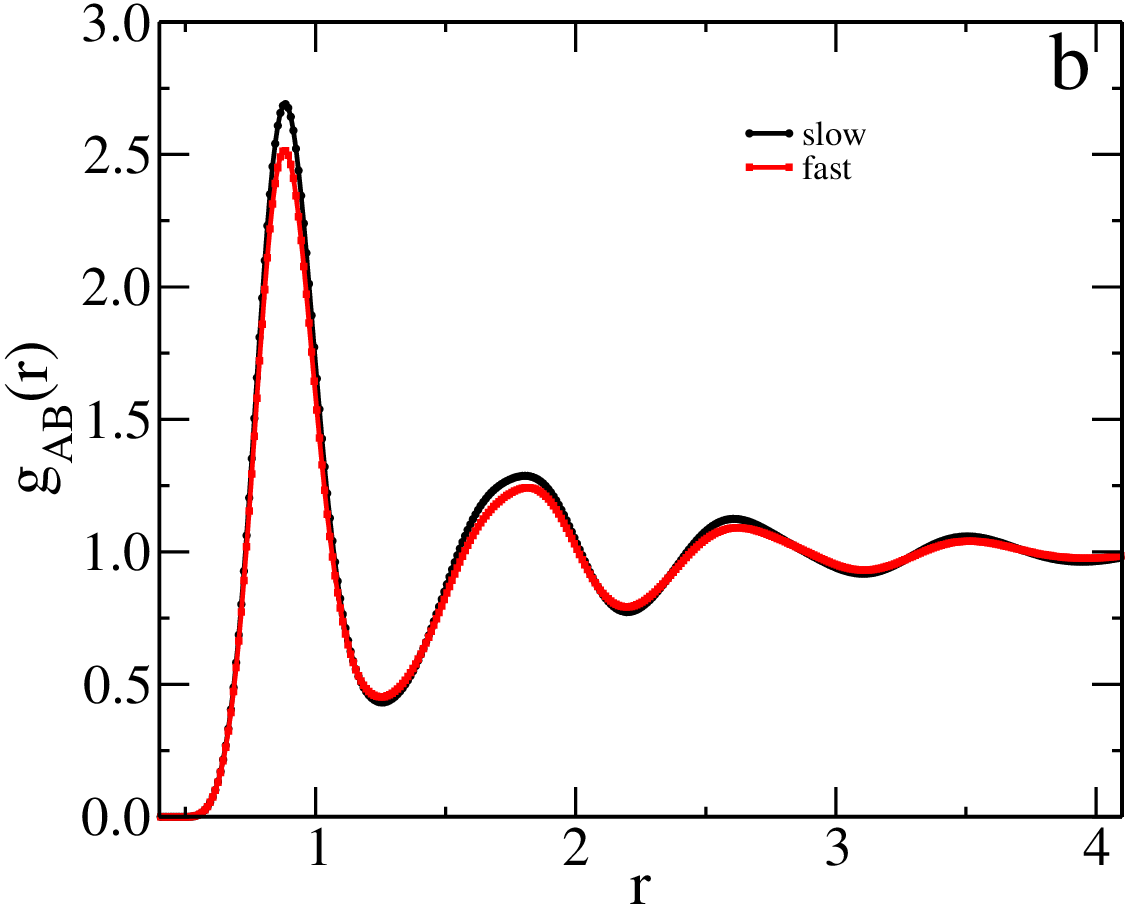}
\end{subfigure}
\caption{The r dependence of $g_{AA}(r)$ and $g_{AB}(r)$ for all particle and it's comparison with the ones which are rearranging. We see that particles which are rearranging have less structured neighbourhood.}
\label{grfast_slow}
\end{figure}

In Fig.\ref{grfast_slow} we plot the rdf of the fast A particles, $g_{AA}(r)$ and $g_{AB}(r)$.  For comparison, we also plot the rdf of all A particles. We find that the neighbourhood of the fast particles are structurally less ordered.\\

\textbf{APPENDIX B: Calculation of local rdf and its effect on the softness parameter}\\

\begin{figure}[!hbt]
\centering
\includegraphics[width=0.35\textwidth,trim = {0 0cm 0 0.1cm},clip]{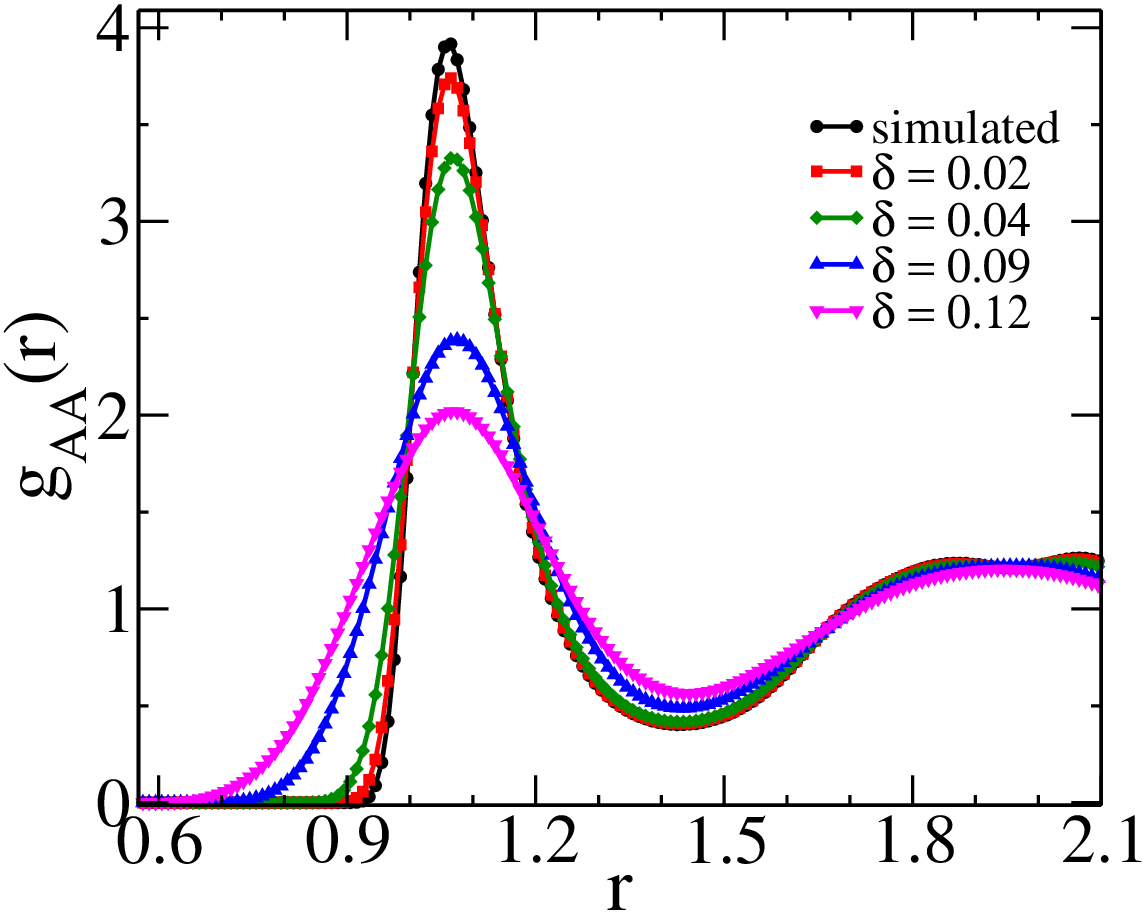}
\caption{$g_{AA}(r)$ vs `r' when the rdf is calculated for each particle using the approximate form given by Eq.\ref{gr} and then averaged over time and particles. For comparison we also plot the simulated rdf. With increase in $\delta$ the first peak becomes wider.}
\label{gr}
\end{figure}

\begin{figure}
\centering
\begin{subfigure}{0.23\textwidth}
\includegraphics[width=0.9\linewidth]{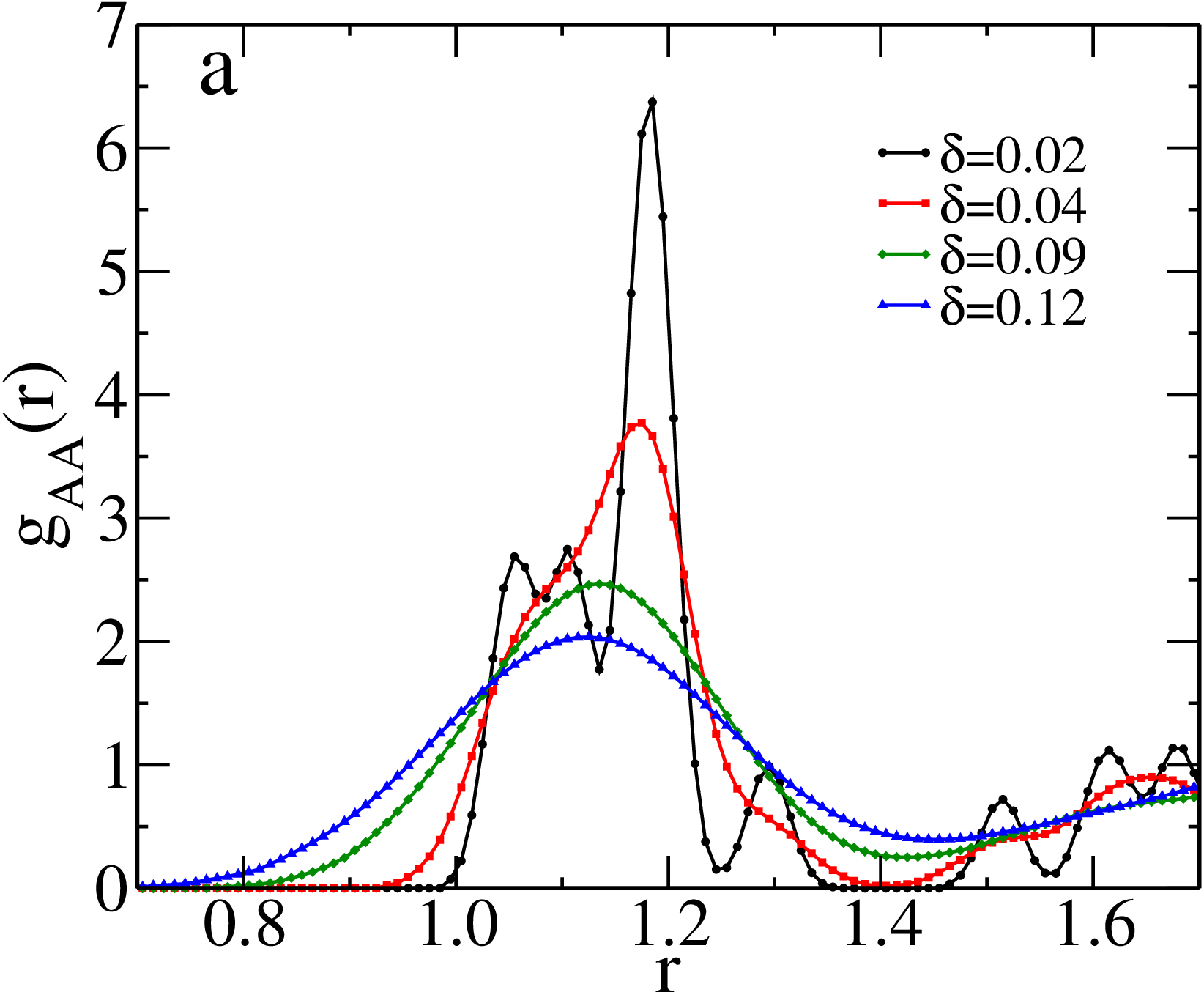}
\end{subfigure}
\begin{subfigure}{0.23\textwidth}
\includegraphics[width=0.9\linewidth]{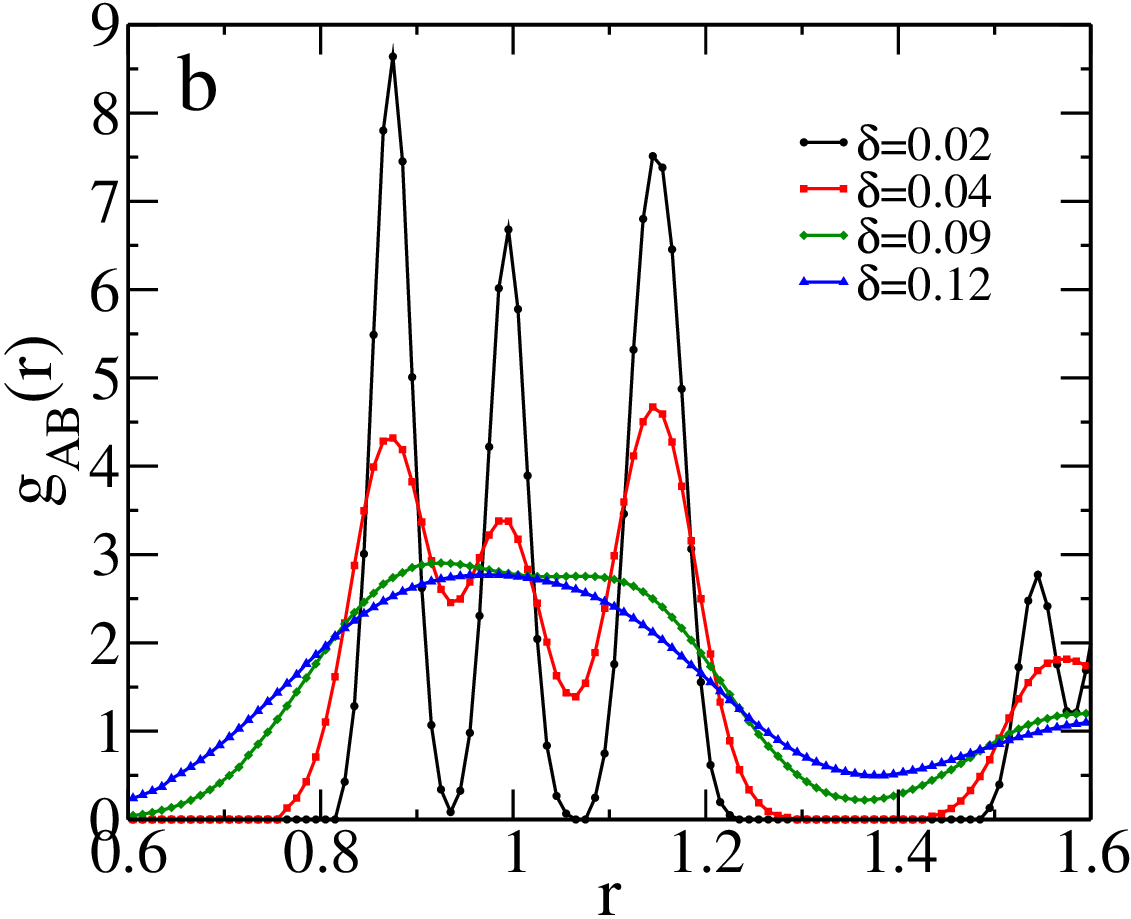}
\end{subfigure}
\caption{$g_{AA}(r)$ $vs$ r for a particle calculated at a single snapshot, for different $\delta$ values. For low $\delta$ values, even the first peak of rdf splits into multiple peaks.}
\label{diff_delta_snapshot}
\end{figure}

\begin{figure}
\centering
\begin{subfigure}{0.23\textwidth}
\includegraphics[width=0.9\linewidth]{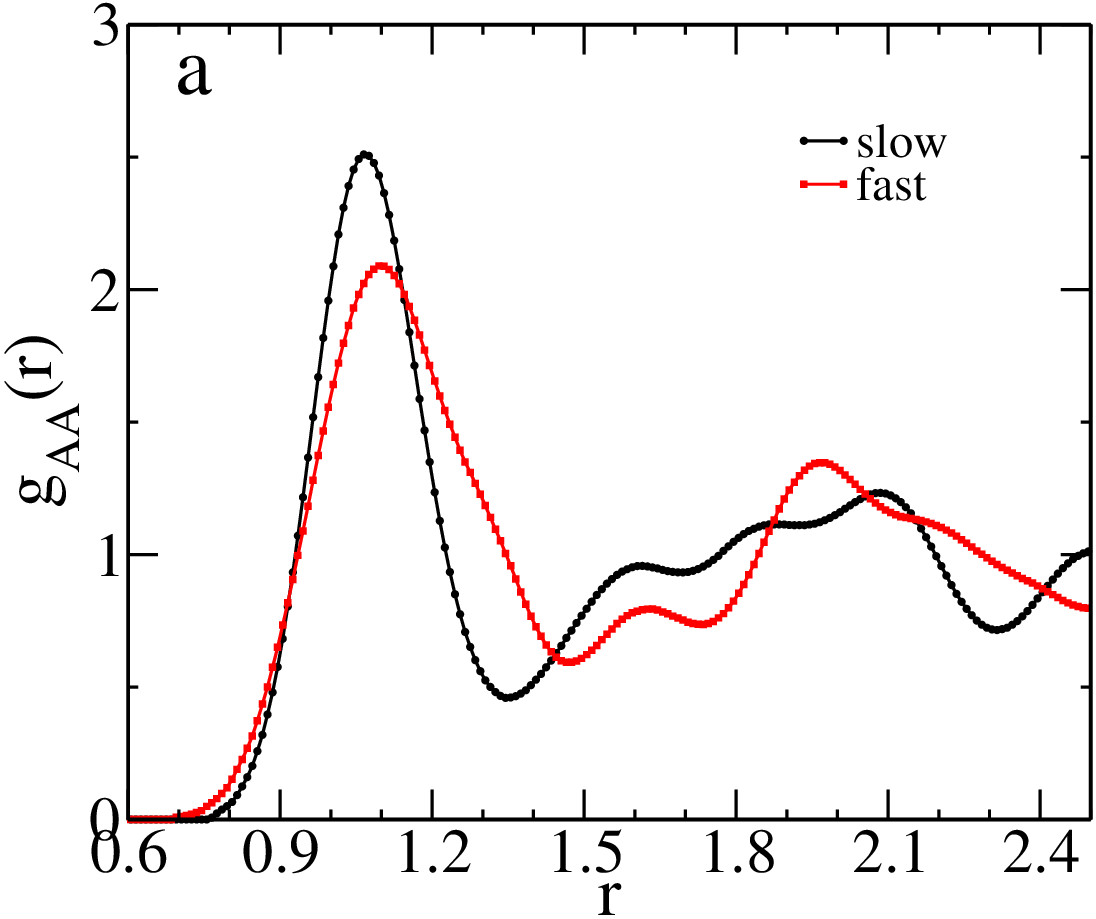}
\end{subfigure}
\begin{subfigure}{0.23\textwidth}
\includegraphics[width=0.9\linewidth]{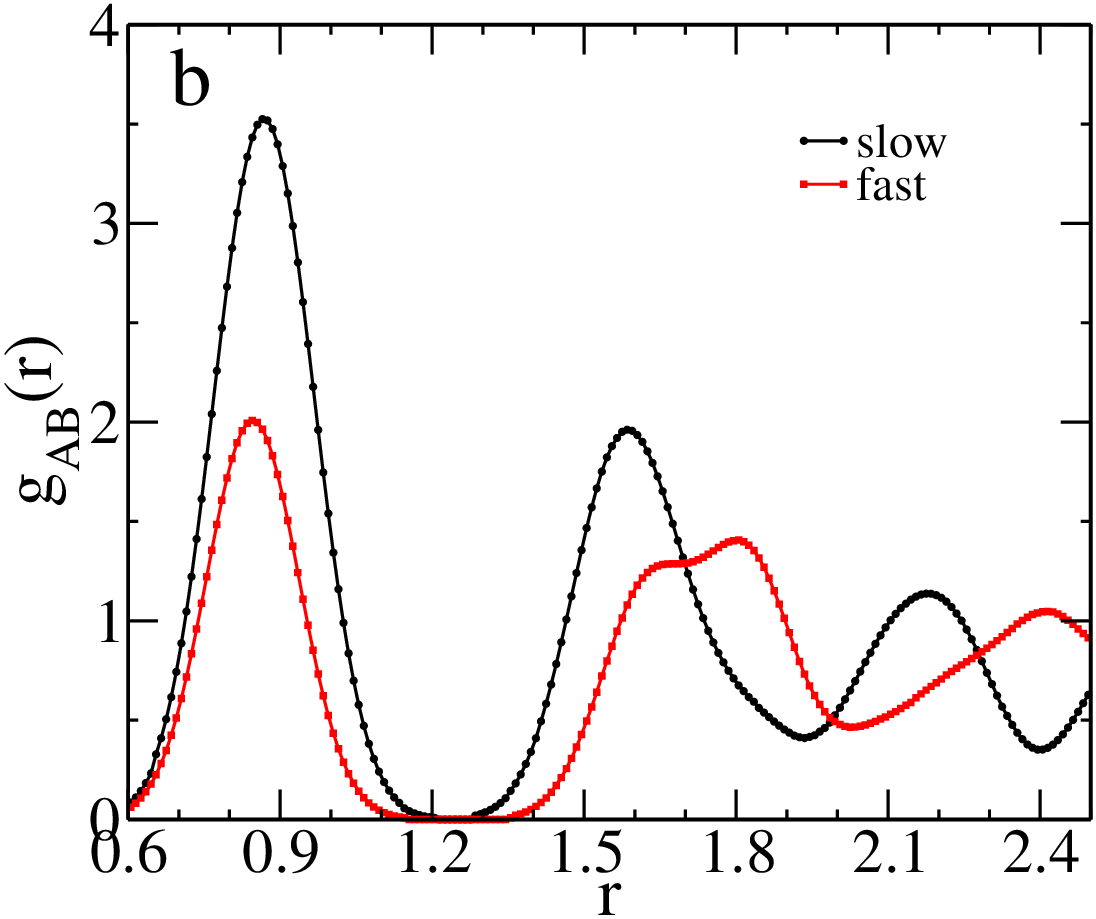}
\end{subfigure}
\caption{Single particle, single snapshot (local) radial distribution function of a representative particle for which $p_{hop}>p_{c}$ (red square)and a particle for which $p_{hop}<p_{c}$ (black circle). The rdf is calculated assuming $\delta=0.09$. It shows that the structure around the fast particle is less well defined.}
\label{particlewise_slow_fast}
\end{figure}

The single particle rdf in single frame can be expressed as a sum of Gaussians and is given by \cite{piaggi}.
\begin{equation}
\begin{aligned}
 g_{\mu\nu}^{i}(r) = \frac{1}{4\pi \rho r^{2}} \sum_{j}\frac{1}{\sqrt{2\pi\delta^{2}}}e^{-\frac{(r - r_{ij})^{2}}{2\delta^{2}}} ,
\end{aligned}
\label{rdf}
\end{equation}
\noindent
where $\delta$ is the variance of Gaussian distribution. The variance is used to make the otherwise discontinuous function a continuous one.  Usually the value of $\delta$ is assumed to be $0.12 \sigma_{BB}$ \cite{piaggi, paddy, richard_prm}. 
Now, if we take an average of $g_{\mu\nu}^{i}$, then we should recover the simulated rdf. 
\begin{equation}
 g_{\mu\nu}(r)=\frac{1}{N}\sum_{i=1}^{N}g_{\mu\nu}^{i}(r) ,
 \label{avgrdf}
\end{equation}
\noindent
where N is the number of particles over which we average the rdf. In Fig.\ref{gr} we plot this average rdf between the A particles for different $\delta$ values. In the same figure, we also plot the simulated rdf, which is obtained in the usual manner by averaging the histograms. As expected we find that with a decrease in the width of the Gaussian (in Eq.\ref{rdf}) the rdf obtained from Eq.\ref{avgrdf} better resembles the average rdf. Thus this introduction of a finite value of $\delta$ does introduce some error in the value of the rdf. In Fig.\ref{diff_delta_snapshot} we plot the rdf obtained  for a particle in a single snapshot for different values of $\delta$ using Eq.\ref{rdf}. We find that as $\delta$ decreases, even the first peak of the rdf splits into multiple peaks. In Fig.\ref{particlewise_slow_fast} we plot the local rdf of a representative particle for which $p_{hop}>p_{c}$ and a particle for which $p_{hop}<p_{c}$. It shows that the structure around the fast particle is less well defined.
\begin{figure}
    \centering
    \includegraphics[width=0.30\textwidth]{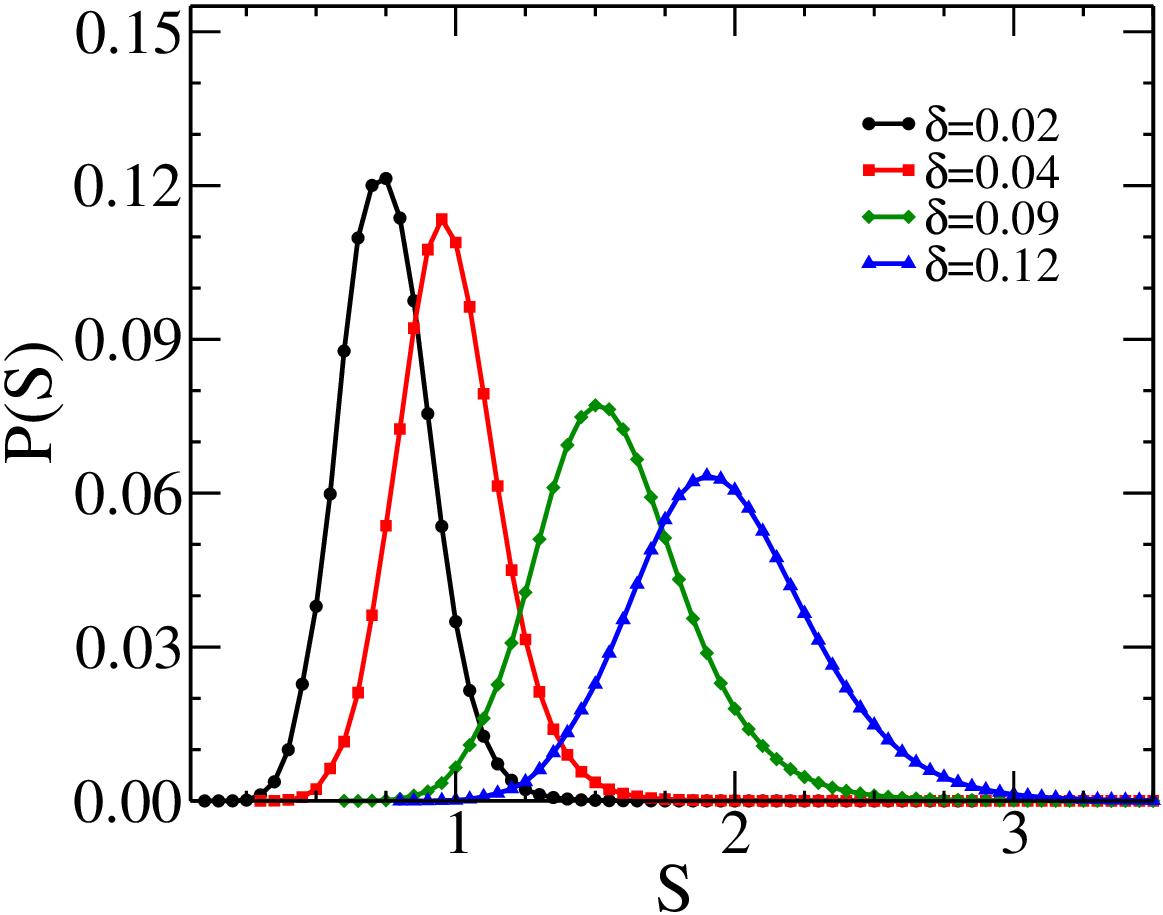}
    \caption{Distribution of softness values for different values of $\delta$ and we see that as $\delta$ increases the corresponding softness distribution also widens this is because the first peak in g(r) also spreads for high $\delta$ values. The softness values mentioned in the axis properties are scaled by $10^{-3}$.}
    \label{softness_dist}
\end{figure}

In the calculation of $\beta\Phi_{r}(\Delta r=0)$ there are two terms, one describing the interaction with other `A' particles and is a function of $g_{AA}$ and the other describing the interaction with the `B' particles and is a function of  $g_{AB}$. We find that the correlation between the dynamics and softness is best described when we set an upper limit to the r integration such that they correspond to the respective minimum of $g_{AA}$ and $g_{AB}$. This implies that it is the first nearest neighbours which contribute most to the softness.
Since we have a finite value of $\delta$, the approximate rdf at low `r' has a finite value where the actual rdf goes to zero. The depth of the mean-field caging potential, $\Phi_{r}(\Delta r=0)$ is calculated as a product of $C(r)$ and g(r) and the $C(r)$ has terms proportional to the interaction potential, $U(r)$ which diverges at small `r'. Thus in this range, small errors in the rdf value get magnified in the calculation of $\Phi_{r}(\Delta r=0)$. To minimize this error, we also put a lower cutoff in the `r' integration. This lower cutoff in `r' range is chosen as the value where the average rdf calculated from the histogram vanishes.

\begin{figure}
\centering
\begin{subfigure}{0.23\textwidth}
\includegraphics[width=0.9\linewidth]{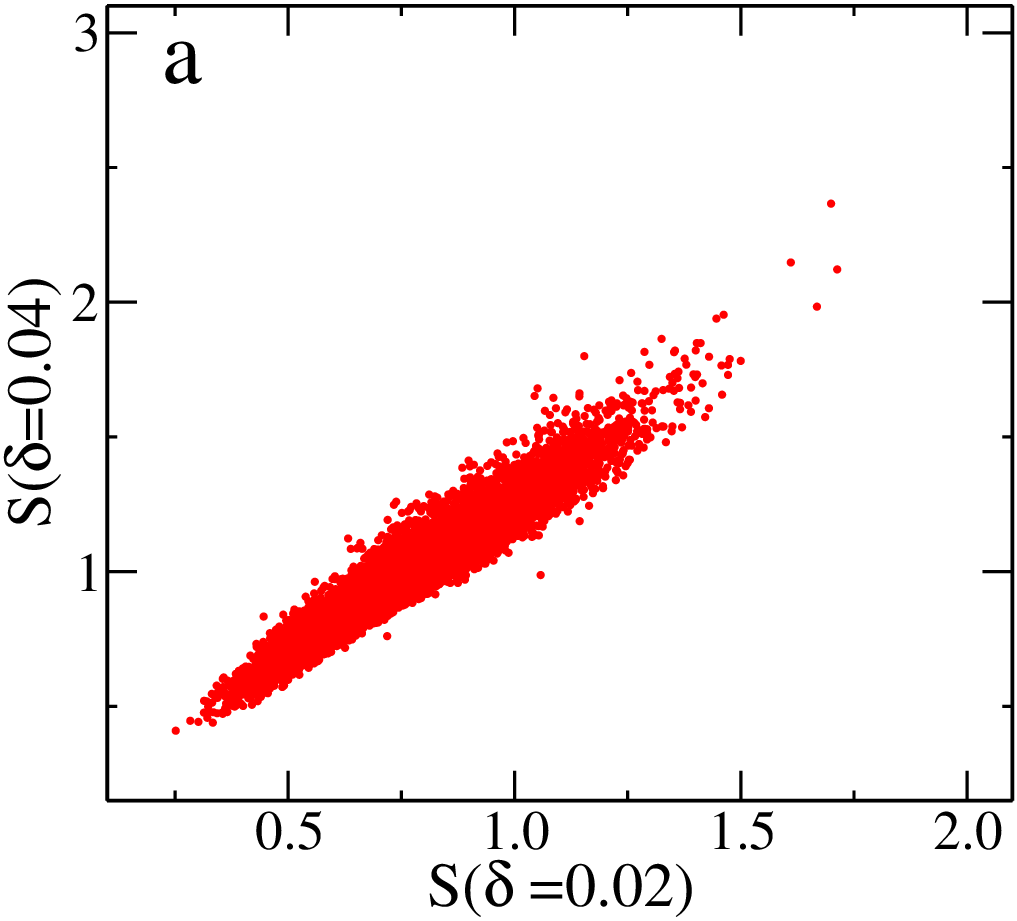}
\end{subfigure}
\begin{subfigure}{0.23\textwidth}
\includegraphics[width=0.9\linewidth]{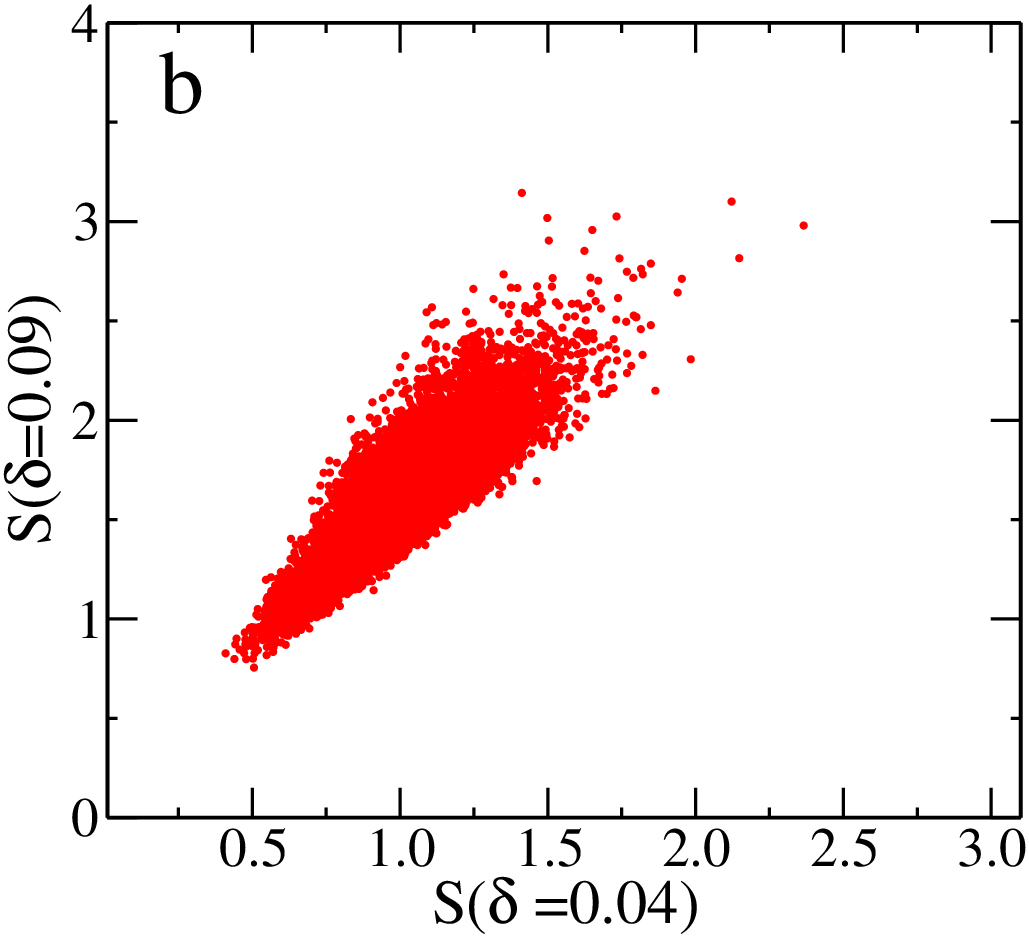}
\end{subfigure}
\vskip\baselineskip
\begin{subfigure}{0.23\textwidth}
\includegraphics[width=0.9\linewidth]{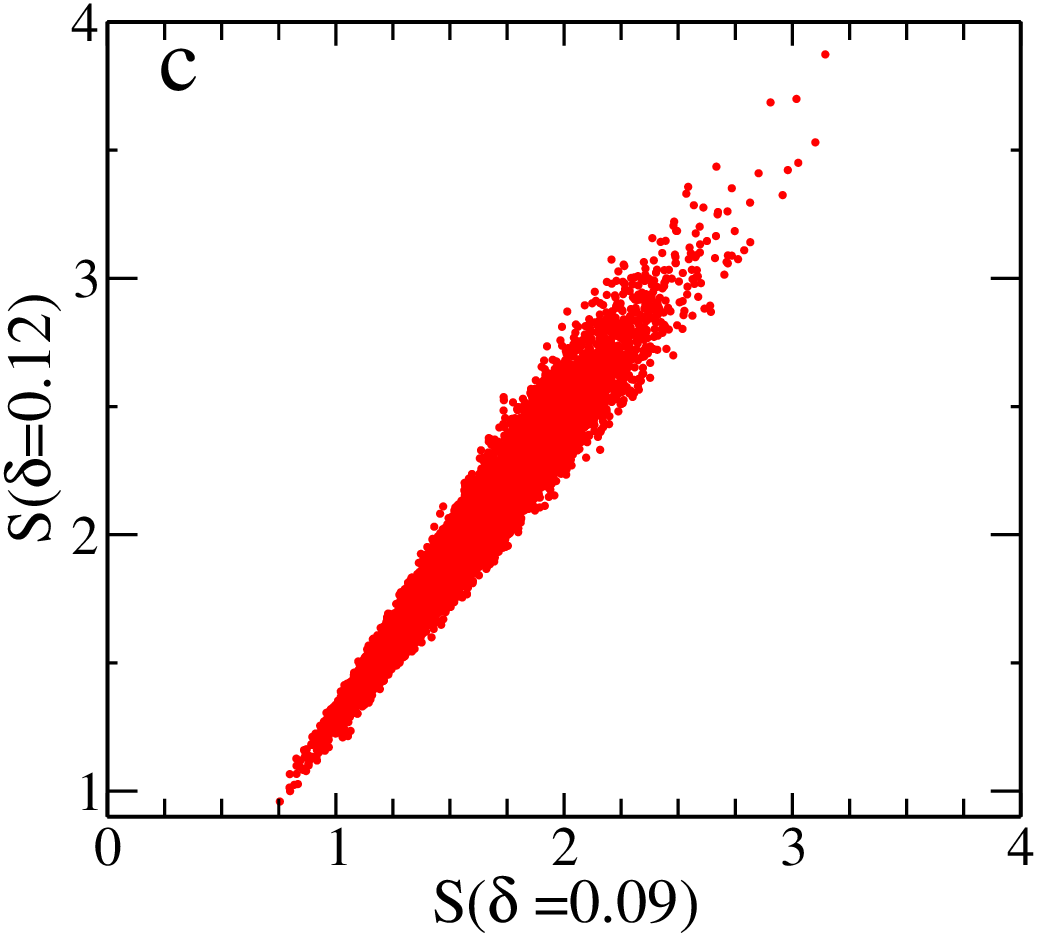}
%,trim = {0 0cm 0 0.1cm},clip]{final_1byS_phiq.eps}
\end{subfigure}
\caption{The correlation of softness at different $\delta$ values.
Although the value of the softness depends on $\delta$ the relative degree of softness of any particle is almost independent of $\delta$. The softness values mentioned in the axis properties are scaled by $10^{-3}$.}
\label{diff_delta_softness}
\end{figure}

As expected, the value of the softness parameter is dependent on the $\delta$ we use to describe the local rdf. However, we show in Fig.\ref{softness_dist} that the variation of $\delta$ just shifts the value of the softness parameter. Since our analysis depends more on the relative value of the softness parameter rather than the exact value, the physics is expected to be independent of $\delta$. In Fig.\ref{diff_delta_softness} we plot the correlation of the softness of the particles calculated using different values of $\delta$. We find that although the value of the softness depends on $\delta$, the relative degree of softness of any particle is almost independent of $\delta$, and the softness values show a strong correlation.

\begin{figure}
\centering
\begin{subfigure}{0.23\textwidth}
\includegraphics[width=0.9\linewidth]{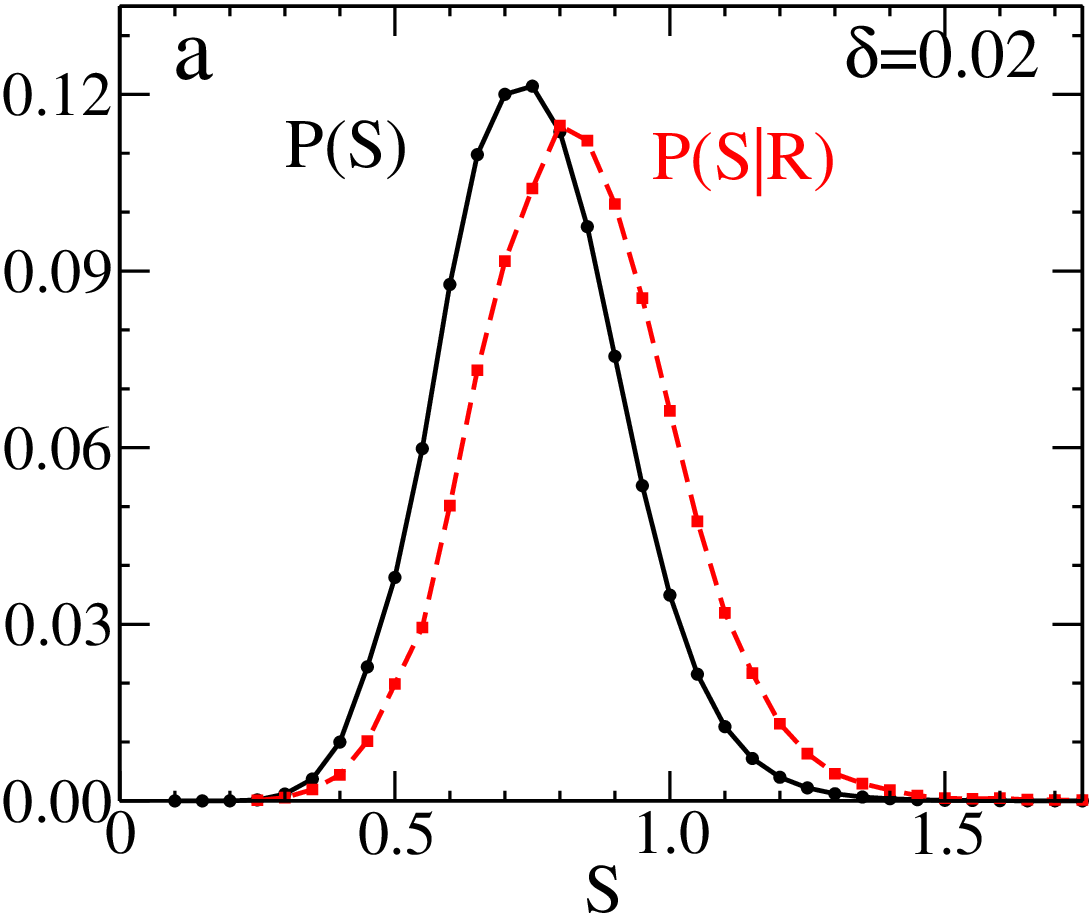}
\end{subfigure}
\begin{subfigure}{0.23\textwidth}
\includegraphics[width=0.9\linewidth]{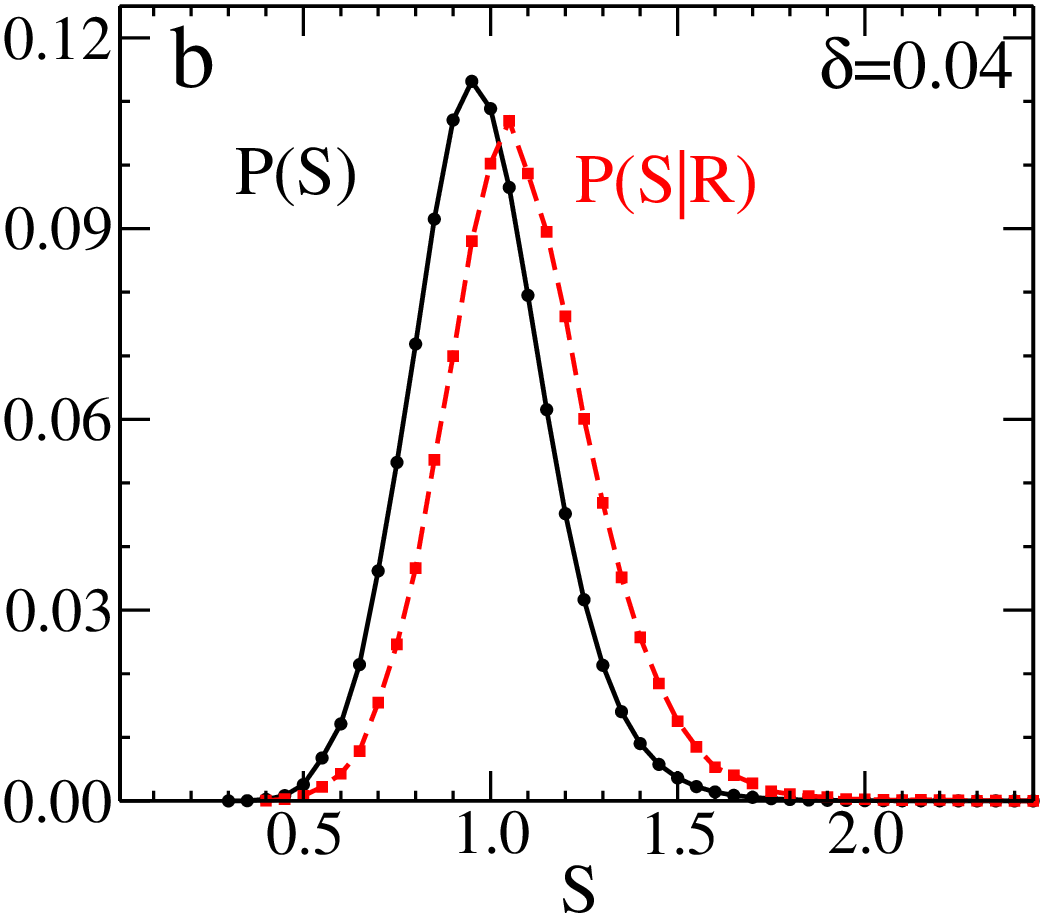}
\end{subfigure}
\vskip\baselineskip
\begin{subfigure}{0.23\textwidth}
\includegraphics[width=0.9\linewidth]{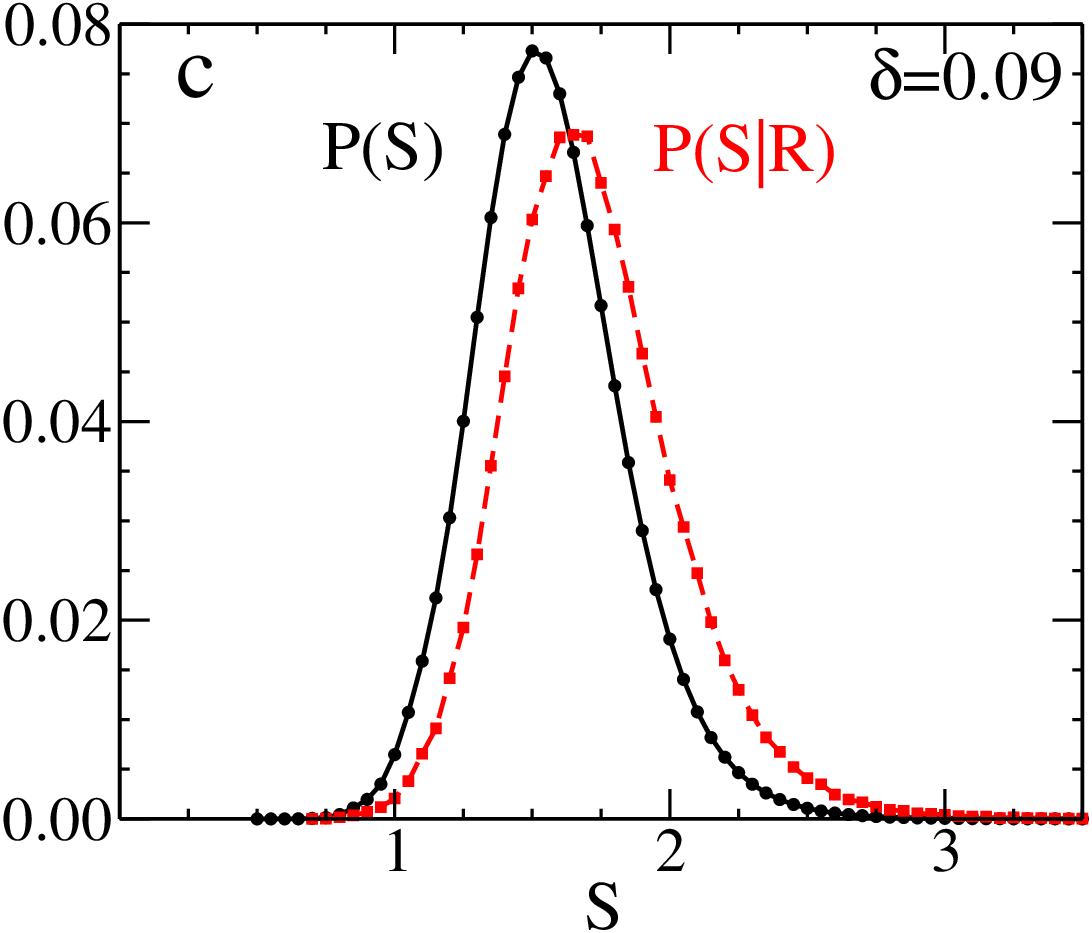}
%,trim = {0 0cm 0 0.1cm},clip]{final_1byS_phiq.eps}
\end{subfigure}
\begin{subfigure}{0.23\textwidth}
\includegraphics[width=0.9\linewidth]{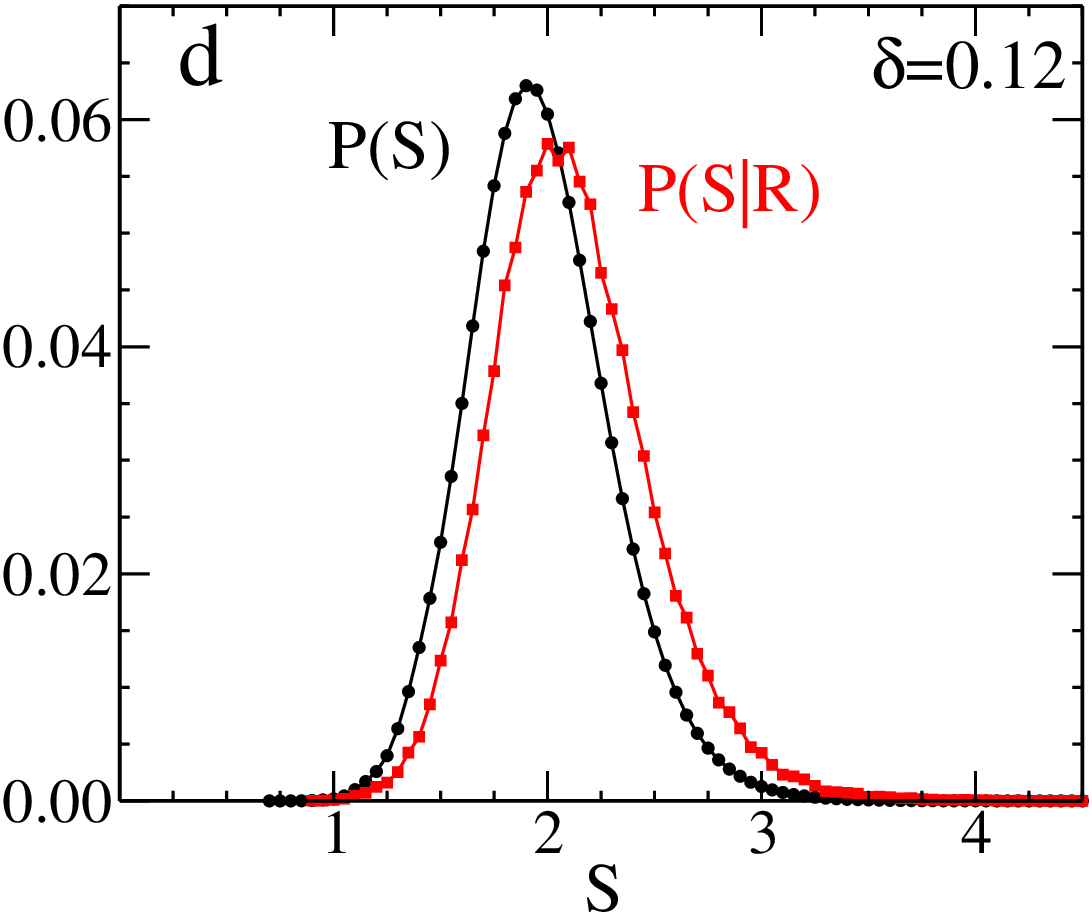}
\end{subfigure}
\caption{The distribution of softness of all particles, $P(S)$ in the system(black) and of those which are about to rearrange(red), $P(S|R)$ for different $\delta$ values. Irrespective of the choice of $\delta$, the softness values of particles which are rearranging are higher. The softness values mentioned in the axis properties are scaled by $10^{-3}$.}
\label{diff_delta_fast}
\end{figure}

In Fig.\ref{diff_delta_fast} we plot the probability distribution of the local softness, $P(S)$ at T=0.47. In the same figure, we also plot the probability distribution of the softness of the particles just before they undergo a rearrangement, $P(S|R)$. This comparison is done for different $\delta$ values and the percentage of particles that undergo rearrangement and has a value of softness greater than the peak varies a little with $\delta$. We find that for $\delta$ = 0.02 it is 66\%, $\delta$ = 0.04 it is 74\%, $\delta$ = 0.09 it is 75.2\%, and for $\delta$ = 0.12 it is 71\%. This is similar to that observed in the Machine learning study where they found that the rdf provides 77\% prediction accuracy of rearrangements \cite{liu_nature}.  

\begin{figure}
\centering
\begin{subfigure}{0.23\textwidth}
\includegraphics[width=0.9\linewidth]{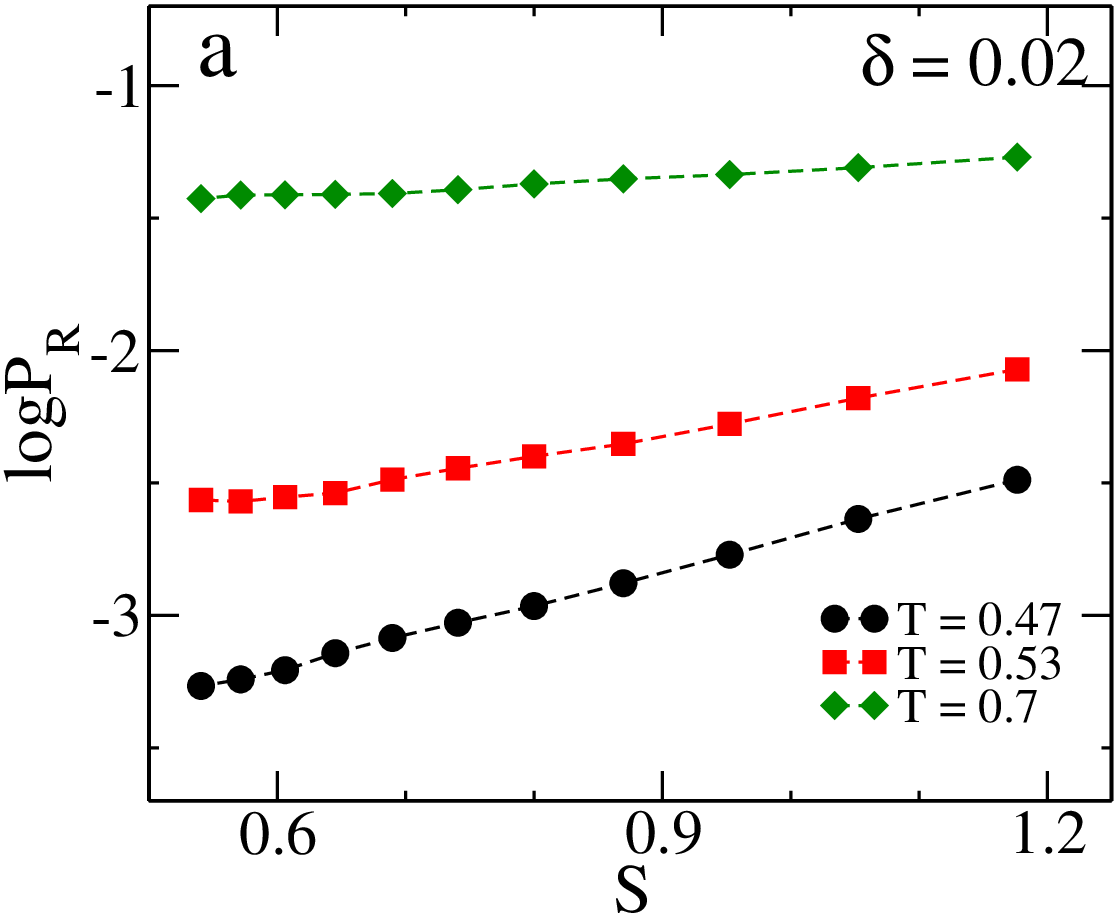}
\end{subfigure}
\begin{subfigure}{0.23\textwidth}
\includegraphics[width=0.9\linewidth]{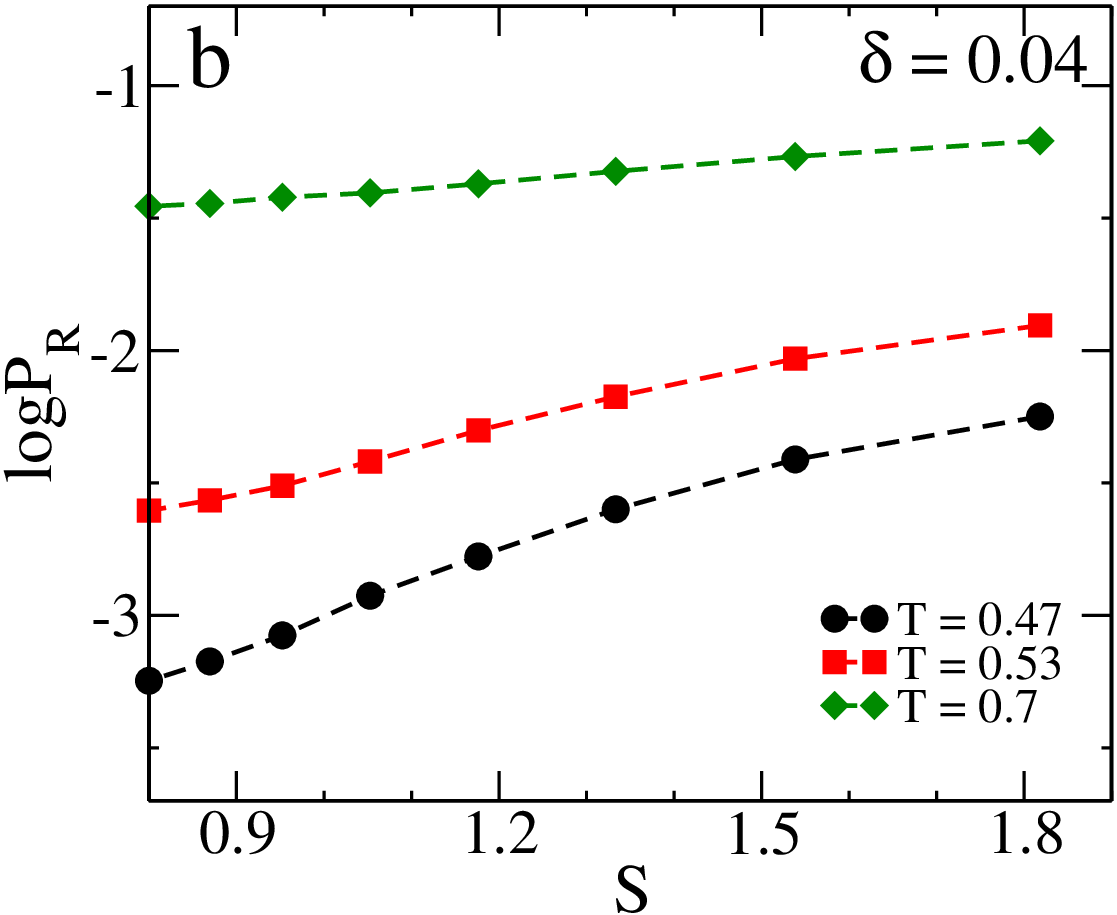}
\end{subfigure}
\vskip\baselineskip
\begin{subfigure}{0.23\textwidth}
\includegraphics[width=0.9\linewidth]{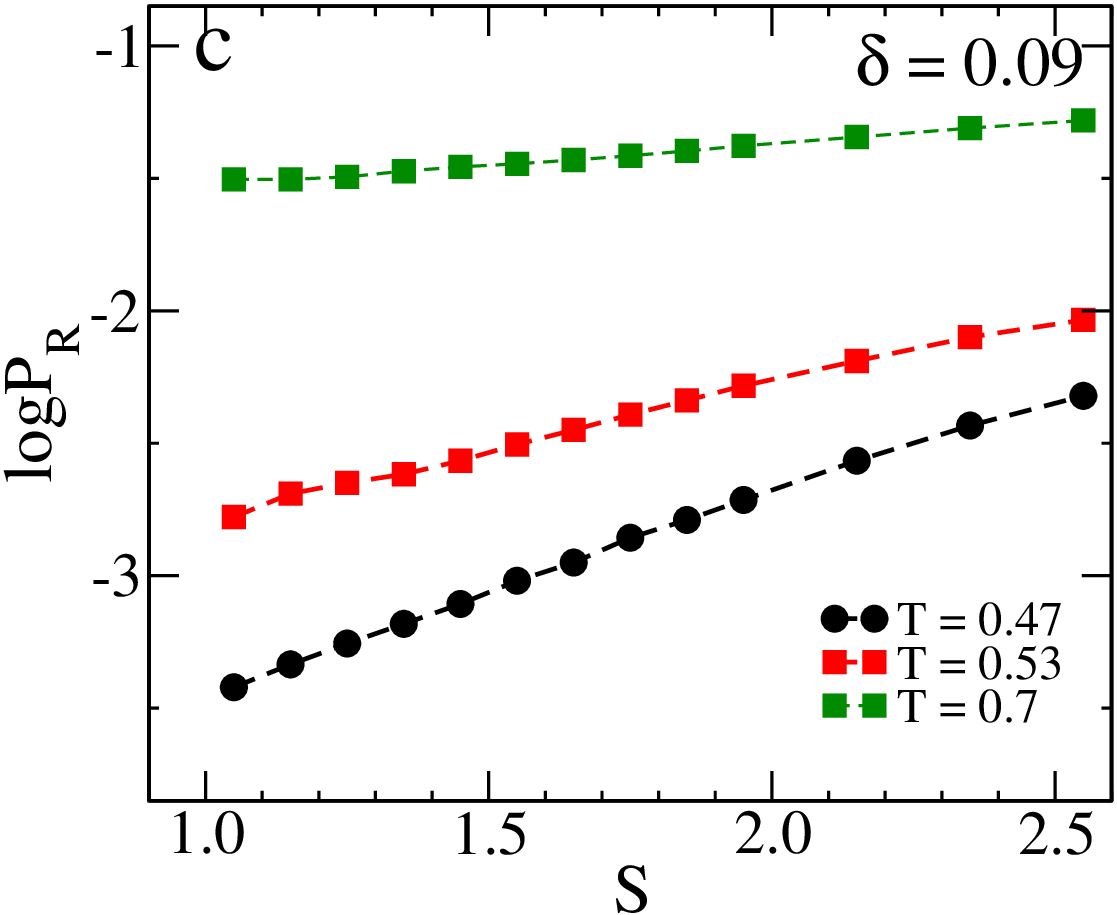}
%,trim = {0 0cm 0 0.1cm},clip]{final_1byS_phiq.eps}
\end{subfigure}
\begin{subfigure}{0.23\textwidth}
\includegraphics[width=0.9\linewidth]{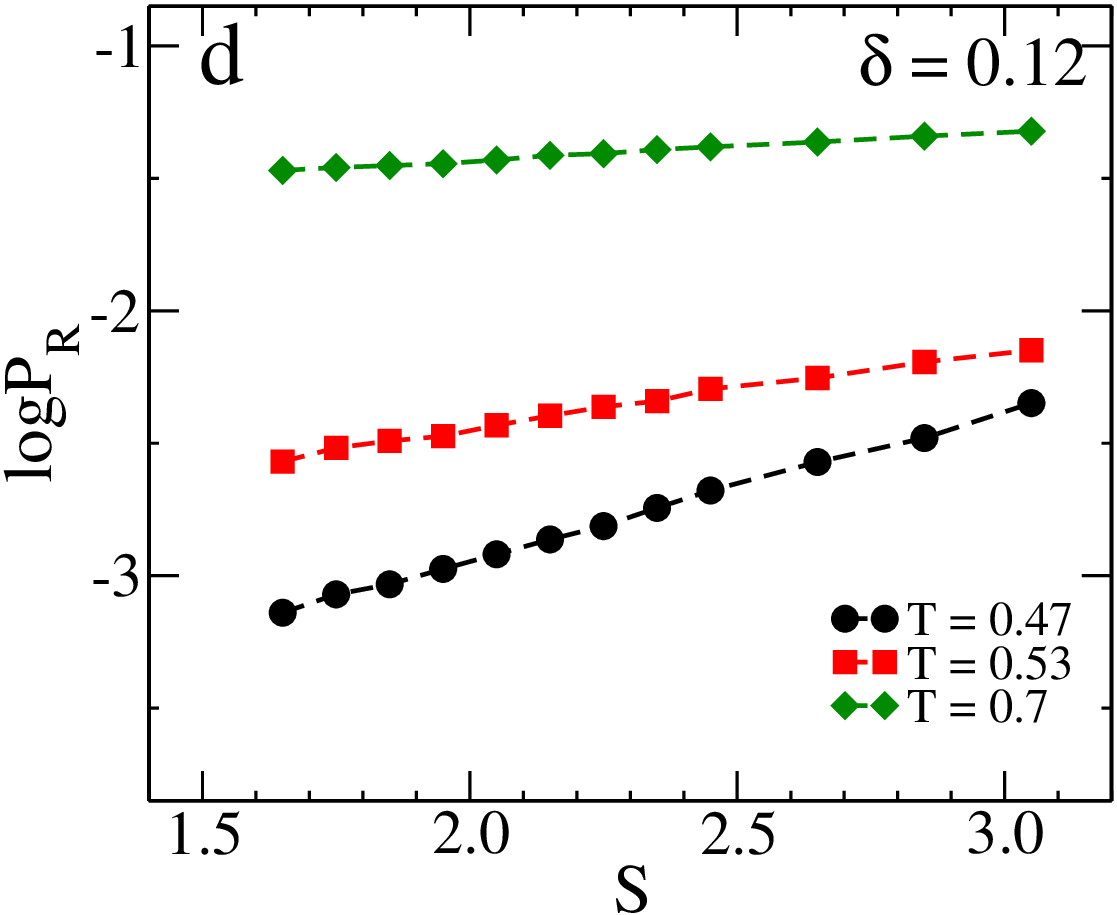}
\end{subfigure}
\caption{$P_{R}(S)$ vs S for different values of $\delta$ at different temperatures. The softness values mentioned in the axis properties are scaled by $10^{-3}$. The
base of logarithm is 10.}
\label{diff_delta_pr}
\end{figure}

\begin{figure}
\centering
\begin{subfigure}{0.23\textwidth}
\includegraphics[width=0.9\linewidth]{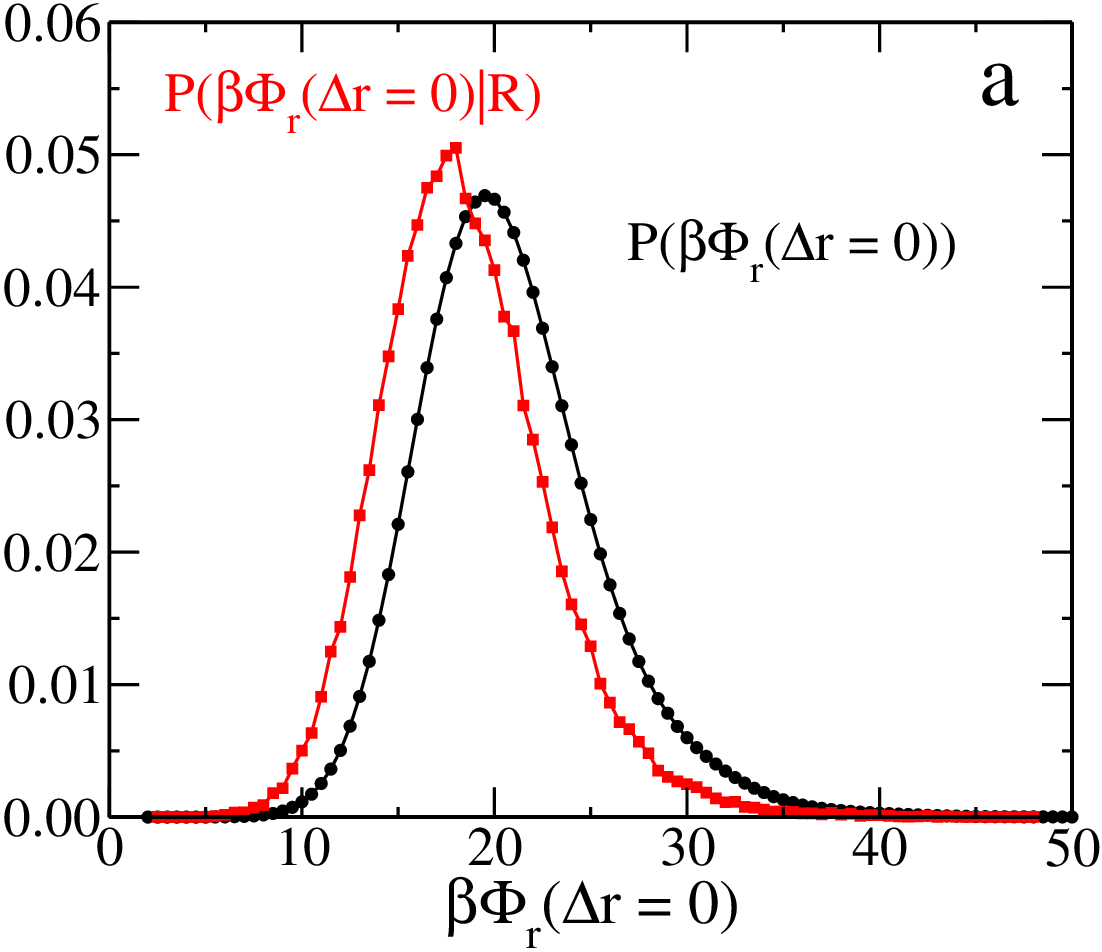}
\end{subfigure}
\begin{subfigure}{0.23\textwidth}
\includegraphics[width=0.9\linewidth]{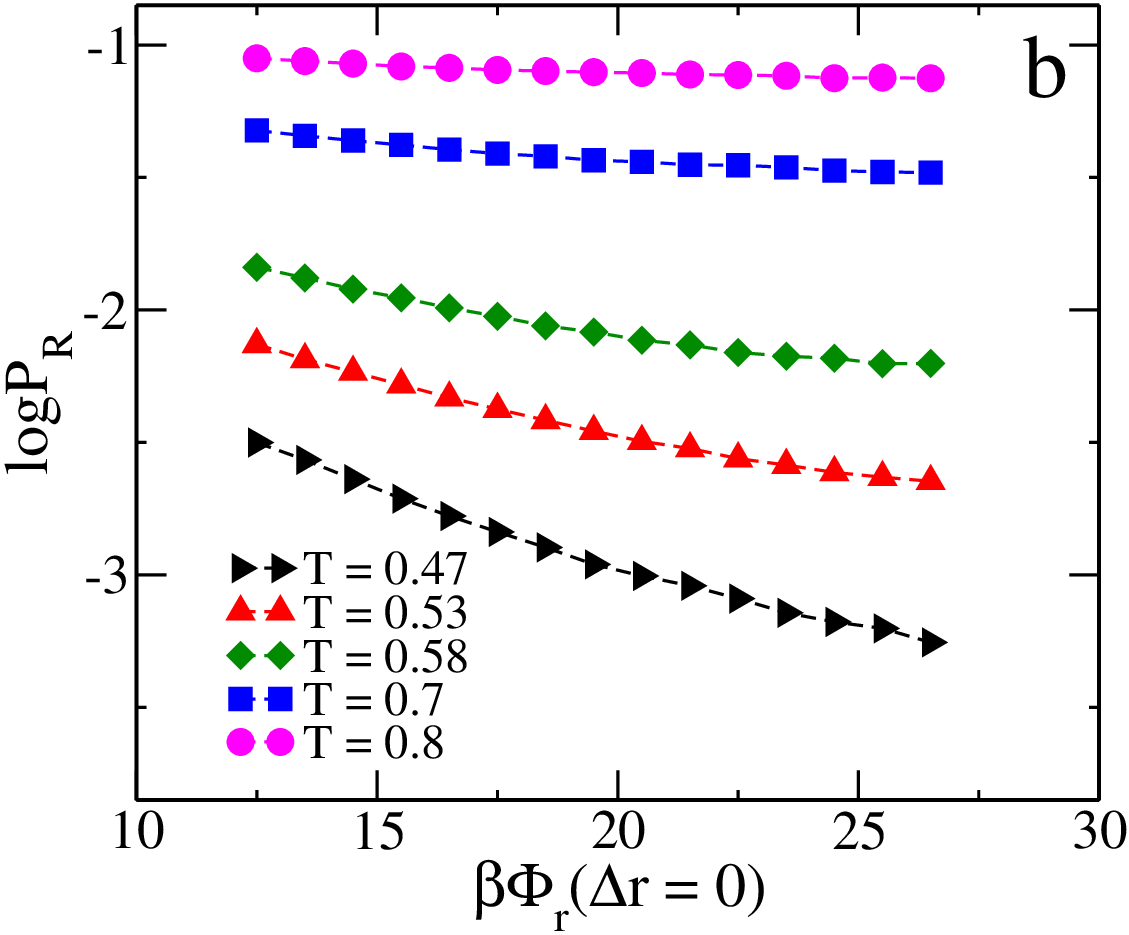}
\end{subfigure}
\vskip\baselineskip
\begin{subfigure}{0.23\textwidth}
\includegraphics[width=0.9\linewidth]{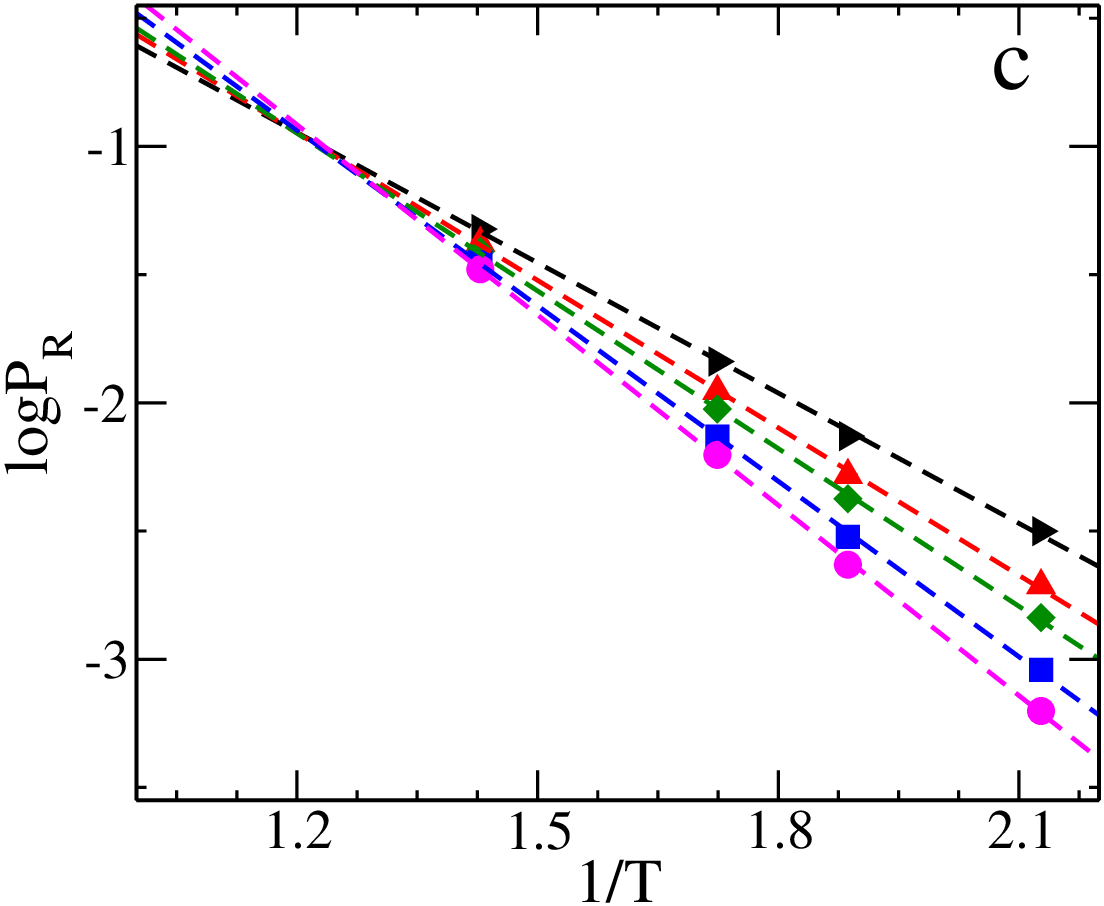}
\end{subfigure}
\begin{subfigure}{0.23\textwidth}
\includegraphics[width=0.9\linewidth]{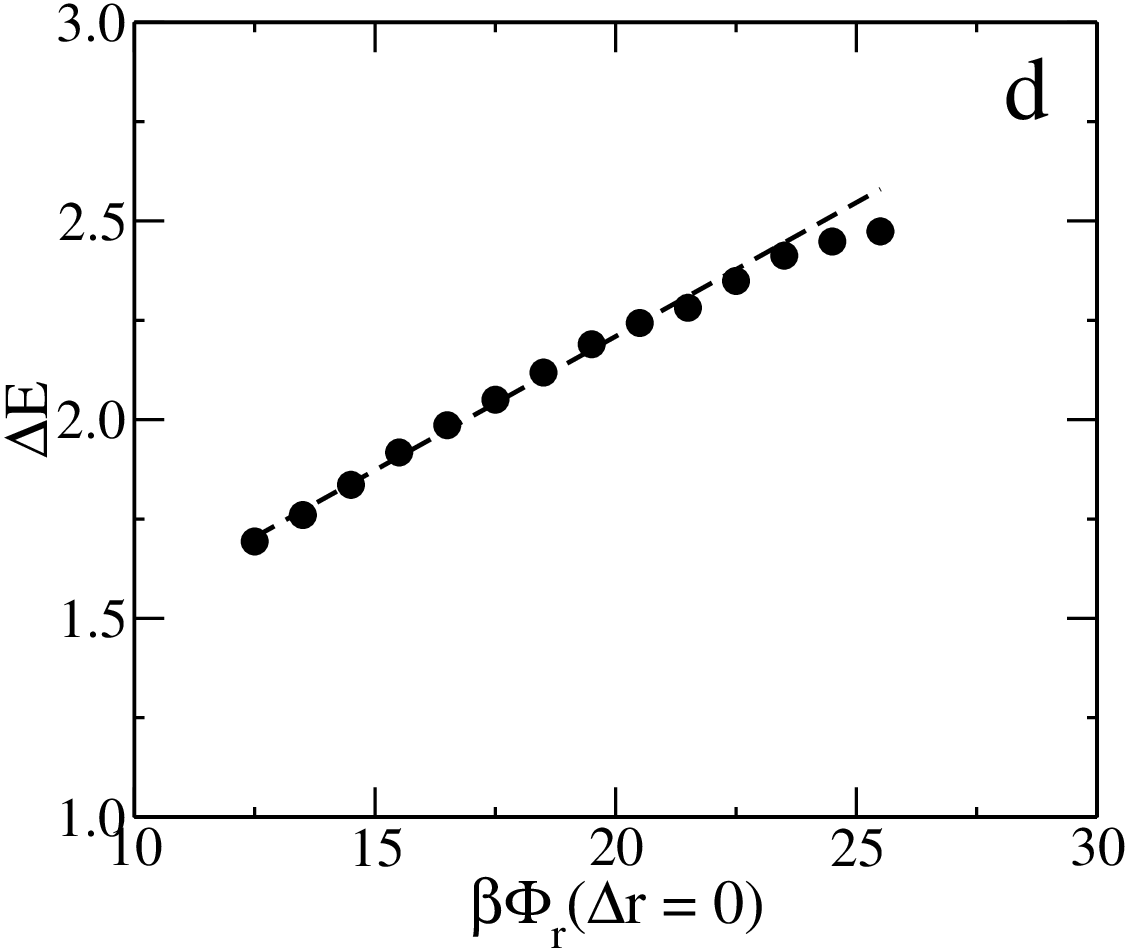}
%,trim = {0 0cm 0 0.1cm},clip]{final_1byS_phiq.eps}
\end{subfigure}
\caption{(a)The distribution of $\beta\Phi_{r}(\Delta r=0)$ for all particle in the system(black) and of those which are about to rearrange(red). (b)Fraction of particle with $\beta\Phi_{r}(\Delta r=0)$ value that undergo rearrangement as a function of $\beta\Phi_{r}(\Delta r=0)$ plotted for different temperatures. (c) $P_{R}$($\beta\Phi_{r}(\Delta r=0)$) as a function of 1/T. (d) $\beta\Phi_{r}(\Delta r=0)$ dependence of energy barrier obtained from (c).The
base of logarithm is 10.}
\label{local_potential}
\end{figure}
In Fig.\ref{diff_delta_pr} we plot the fraction of particles that undergo rearrangement as a function of softness at few temperatures. This plot is done at different $\delta$ values. We find that irrespective of the $\delta$ value, the study shows that the probability of rearrangement depends on softness, and this dependence is stronger at lower temperatures. For the rest of the study, we assume $\delta=0.09$, which is similar to the value used in earlier studies \cite{piaggi,paddy}.

In Fig.\ref{local_potential} we study the causal relationship between the depth of the local caging potential $\beta \Phi_{r}(\Delta r=0)$ and the local rearrangements. We find that 62$\%$ of the particles that undergo rearrangement has a depth of potential lower than the peak value of the distribution (Fig.\ref{local_potential}(a)). We also find that the probability of rearrangement is dependent on $\Phi_{r}(\Delta r=0)$ and this dependence becomes stronger at lower temperatures (Fig.\ref{local_potential}(b)). We also find that the dynamics can be expressed in an Arrhenius form where the barrier is dependent on $\Phi_{r}(\Delta r=0)$ (Fig.\ref{local_potential}(c)). The correlation between the local rearrangement and the depth of the local potential is present only below the temperature where the $log_{10}P_{R}$ vs $1/T$ plots cross each other. This temperature T=0.80 is similar to that predicted from the softness parameter and is connected to the onset temperature of the system \cite{atreyee_onset}. The barrier for the Arrhenius dynamics is linearly proportional to the depth of the caging potential. \\

%{\bf Availability of Data}\\
%The data that support the findings of this study are available from the corresponding author upon reasonable request.\\

{\bf References}\\
\bibliographystyle{h-physrev}
\bibliography{final_reference.bib}

\end{document}